\documentclass[preprint,12pt]{elsarticle}

\usepackage{amsmath}
\usepackage{amssymb}
\usepackage{amsthm}

\usepackage{lmodern}

\usepackage{enumitem}

\usepackage{tikz}
\usetikzlibrary{arrows.meta,calc,positioning,fit,backgrounds}

\theoremstyle{plain}
\newtheorem{theorem}{Theorem}[section]
\newtheorem*{theorem*}{Theorem}
\newtheorem{lemma}{Lemma}[section]
\newtheorem{proposition}{Proposition}[section]
\newtheorem{corollary}{Corollary}[section]
\theoremstyle{definition}
\newtheorem{definition}{Definition}[section]
\newtheorem*{definition*}{Definition}
\newtheorem{example}{Example}[section]

\theoremstyle{remark}

\journal{Journal of Economic Theory}

\begin{document}

\begin{frontmatter}

\title{Stability and Perfection in Constrained Weighted Network Games}

\author[mcgt]{Julien Fixary}
\author[mcgt]{Rémi Castera}
\author[mcgt]{Rida Laraki}

\affiliation[mcgt]{
    organization={Moroccan Center for Game Theory},
    addressline={Université Mohammed VI Polytechnique}, 
    city={Rabat},
    country={Morocco}
    }

\begin{abstract}
    We introduce a general framework of constrained weighted network formation in which agents allocate resources across coalitional interactions while facing feasibility constraints.
    More precisely, we allow arbitrary coalitions of agents to be assigned a continuous weight, and we incorporate both coalitional capacity constraints and individual budget constraints into the model.
    This framework extends the weighted network formation model of Bich and Morhaim \cite{bich_morhaim_2020} by combining continuous interaction intensities with endogenous feasibility restrictions.
    Within this setting, we introduce a new notion of stability -- called \emph{coalitionwise stability} -- extending pairwise stability to constrained network games.
    Under standard continuity and componentwise quasiconcavity assumptions on payoff functions, we establish the existence of coalitionwise stable networks.
    We then investigate refinements of stability inspired by the notion of perfect equilibrium introduced by Selten \cite{selten_1975}, and by the notion of perfect pairwise stability developed by Bich and Teteryatnikova \cite{bich_teteryatnikova_2023} for network games.
    The presence of feasibility constraints fundamentally alters the geometry of the feasible set, which no longer possesses a product structure.
    As a consequence, trembling-hand type refinements become substantially more delicate, since perturbations of some interactions may render other interactions infeasible.
    Nevertheless, we introduce constrained analogues of both the trembling-hand and perturbative notions of perfect stability considered by Bich and Teteryatnikova \cite{bich_teteryatnikova_2023}, and study their existence and their relationship. In particular, we show the equivalence between these trembling-hand and perturbative formulations under some componentwise monotonicity assumption.
\end{abstract}

\begin{highlights}
    \item We introduce a general framework of constrained weighted network formation with coalitional capacity constraints and individual budget constraints.
    \item We define \emph{coalitionwise stability}, extending pairwise stability to constrained weighted network games.
    \item We prove the existence of coalitionwise stable networks under continuity and componentwise quasiconcavity.
    \item We introduce constrained trembling-hand and perturbative refinements of stability and establish their existence.
    \item We characterize the relationship between trembling-hand and perturbative notions of perfection under feasibility constraints.
\end{highlights}


\end{frontmatter}

\section{Introduction}

Following the seminal contribution of Jackson and Wolinsky \cite{jackson_wolinsky_1996}, network formation theory has become a major field of research in microeconomic theory.
Its objective is to understand how networks emerge endogenously from the strategic interactions of self-interested agents, and how individual incentives shape the architecture of social, economic, or informational structures.
Over the last decades, the literature has developed a wide variety of models and solution concepts in order to analyze the formation, evolution, efficiency, and robustness of networks (see, e.g., \cite{bich_morhaim_2020,bich_teteryatnikova_2023,calvo-armengol_ilkilic_2009,herings_mauleon_vannetelbosch_2009,jackson_van-den-nouweland_2005}).
Beyond its theoretical importance, network formation theory has also found numerous applications in economics and the social sciences, including the study of educational interactions \cite{calvo-armengol_patacchini_zenou_2009}, urban interactions \cite{helsey_zenou_2014}, R\&D collaboration networks \cite{konig_liu_zenou_2019}, and delinquency and conformism \cite{patacchini_zenou_2009}.

Among the various stability notions introduced in this literature, \emph{pairwise stability}, proposed by Jackson and Wolinsky \cite{jackson_wolinsky_1996}, has become one of the most influential.
In their original framework, networks are unweighted: each pair of agents is either connected or disconnected. A network is said to be pairwise stable if no agent wishes to unilaterally sever one of their existing links, and if no pair of agents can jointly create a new link that would benefit both of them.
This notion captures a local and decentralized form of stability that is particularly well suited to environments in which links are formed bilaterally and require mutual consent.

A substantial part of the subsequent literature has focused on the existence and properties of pairwise stable networks.
Jackson and Watts \cite{jackson_watts_2001}, for instance, show that pairwise stable networks do not necessarily exist in general environments, since local improving paths may cycle indefinitely.
Their work investigates the structure of such improving dynamics and identifies conditions under which pairwise stable networks can nevertheless be reached through local adjustments. In parallel, survey contributions such as \cite{demange_wooders_2005} provide a systematic overview of network formation models and stability concepts, emphasizing the behavioral interpretation of pairwise stability as well as its relationship with efficiency and welfare considerations.

More recently, the literature has progressively moved beyond binary interactions toward weighted network models. In many economic and social applications, relationships are not merely present or absent, but vary in intensity, frequency, or capacity. Motivated by this observation, Bich and Morhaim \cite{bich_morhaim_2020} introduced a framework of weighted network formation in which the intensity of each bilateral interaction is chosen strategically and endogenously by the two agents involved in the relationship. Such weighted interactions naturally arise in a broad variety of environments. In social or communication networks, weights may represent attention, interaction frequency, or time allocation. In financial networks, they may capture exposure sizes or lending intensities. In labor markets, they may describe the strength of mentoring or referral relationships, while in production or R\&D networks they may correspond to collaboration efforts or trade volumes.

Within this weighted framework, pairwise stability admits a natural continuous counterpart. A weighted network is pairwise stable if no agent wishes to reduce the intensity of one of their existing links, and if no pair of agents can jointly increase the intensity of their mutual relationship in a way that benefits both of them.
One of the major contributions of Bich and Morhaim \cite{bich_morhaim_2020} is to establish existence results for pairwise stable weighted networks under broad assumptions on preferences, namely continuity and componentwise quasiconcavity of payoff functions. These results considerably broaden the scope of network formation theory and provide a flexible framework capable of encompassing many economically relevant situations.

Beyond existence questions, recent contributions have also investigated the global and generic properties of pairwise stable weighted networks.
Bich and Fixary \cite{bich_fixary_2022}, for example, study the topological structure of graphs of pairwise stable weighted networks and show that, for broad classes of network games, the number of pairwise stable weighted networks is generically odd.
This oddness property reveals deep connections between pairwise stability and classical equilibrium concepts from game theory and general equilibrium theory.
Building on this perspective, Fixary \cite{fixary_2025} further investigates the global geometry of graphs of pairwise stable weighted networks and establishes, in several important settings, unknottedness and related topological properties shedding light on equilibrium selection and network dynamics.
Computational aspects have also recently attracted increasing attention.
Herings and Zhan \cite{herings_zhan_2024}, for instance, develop homotopy-based procedures to compute pairwise stable weighted networks, while Chen, Tao, and Zhan \cite{chen_tao_zhan_2025} propose efficient path-following algorithms for their computation and selection.

Despite the flexibility of the weighted network formation model, most of the existing literature still abstracts from an important feature of many economic and social environments: agents typically face global constraints on the allocation of their resources across relationships. In practice, individuals possess limited amounts of time, effort, attention, capital, or organizational capacity, which must be distributed across their various interactions. Likewise, groups or coalitions often face exogenous limits on the intensity of the activities they can jointly undertake. Such considerations naturally lead to the introduction of feasibility constraints into network formation models.

The first objective of the present paper is precisely to incorporate these aspects into the theory of weighted network formation. More specifically, we extend the existing framework in two distinct directions. First, instead of restricting attention to bilateral interactions, we allow arbitrary coalitions of agents to be assigned a weight. Under a time-sharing interpretation, this means that agents may allocate resources not only to pairwise relationships, but also to interactions involving larger groups. Second, we explicitly introduce feasibility constraints into the model, namely coalitional capacity constraints and individual budget constraints. Intuitively, each coalition is subject to a maximal feasible intensity, while each agent disposes of a limited amount of resources that must be allocated across all coalitions in which they participate.

To study stability in this setting, we introduce a notion of stability -- called \emph{coalitionwise stability} -- extending both the pairwise stability concept of Jackson and Wolinsky \cite{jackson_wolinsky_1996} and its weighted counterpart introduced by Bich and Morhaim \cite{bich_morhaim_2020}.
Under assumptions analogous to those used in the Bich--Morhaim model -- namely continuity and componentwise quasiconcavity of payoff functions -- we establish the existence of coalitionwise stable weighted networks in constrained network games (see Theorem \ref{thm_1}).

The second objective of this paper concerns refinements of stability and robustness to perturbations. In finite normal-form games, the notion of Nash equilibrium is known to admit equilibria sustained by weakly dominated behaviors or by highly non-robust strategic configurations. In his seminal work on equilibrium refinements, Selten \cite{selten_1975} introduced the notion of \emph{perfect equilibrium} in order to eliminate such pathological outcomes. The central idea is that equilibrium behavior should remain optimal even when players may commit arbitrarily small mistakes, or \emph{trembles}. Selten's original definition is perturbative in nature: players are forced to assign a strictly positive probability to every pure strategy, and perfect equilibria arise as limits of equilibria of these perturbed games.

An important aspect of Selten’s contribution is that he subsequently proved that this perturbative definition admits an equivalent characterization in terms of trembling hands. More precisely, a strategy profile is perfect if it remains optimal against nearby completely mixed perturbations of the opponents' strategies. This characterization provides a more behavioral interpretation of perfection and has become one of the standard viewpoints on equilibrium refinements in non-cooperative game theory.

Inspired by Selten's approach, Bich and Teteryatnikova \cite{bich_teteryatnikova_2023} introduced a refinement of pairwise stability for weighted network games, referred to as \emph{perfect pairwise stability}. Their objective is conceptually similar to Selten's original motivation: standard pairwise stable networks may fail to be robust to perturbations of the environment and may rely on fragile local link decisions. In order to refine pairwise stability, Bich and Teteryatnikova \cite{bich_teteryatnikova_2023} transpose Selten's trembling-hand philosophy to the framework of weighted network formation. Roughly speaking, their notion requires that the stability of each link persists under small perturbations of the intensities of the other links.

A particularly interesting aspect of their work is that they subsequently establish an equivalence between this trembling-hand formulation and a perturbative characterization involving stable networks in suitably perturbed environments. Thus, while Selten \cite{selten_1975} originally introduced perfection through perturbations and later characterized it through trembling-hand arguments, the approach of Bich and Teteryatnikova \cite{bich_teteryatnikova_2023} proceeds in the reverse direction: trembling-hand stability is introduced first, and perturbative stability subsequently appears as an equivalent characterization.

The final objective of the present paper is to investigate how these notions of perfect stability can be adapted to constrained network games. The introduction of budget and feasibility constraints fundamentally changes the geometry of the feasible set, which no longer possesses the product structure encountered in classical weighted network games. As a consequence, perturbations of some links may render other link choices infeasible, creating substantial additional difficulties for trembling-hand type refinements (see Proposition \ref{prop_1} and Proposition \ref{prop_2}). Motivated by these issues, we introduce constrained analogues of both notions considered by Bich and Teteryatnikova \cite{bich_teteryatnikova_2023} -- namely, a trembling-hand notion of perfect stability and a perturbative notion based on perturbations of the feasible environment itself. We then investigate their existence (see Theorem \ref{thm_3} and Theorem \ref{thm_5}), and analyze the extent to which equivalence results analogous to those obtained by Selten \cite{selten_1975} and by Bich--Teteryatnikova \cite{bich_teteryatnikova_2023} can still be obtained in the presence of feasibility constraints (see Theorem \ref{thm_4}).

\section{The Model}

As explained in the introduction, there are several limitations to the Bich--Morhaim model: the aim of this section is to extend it along two dimensions.
First, interactions are no longer restricted to bilateral links: weights are assigned to coalitions of agents, allowing us to model multilateral relationships.
Second, agents' choices are subject to constraints, reflecting limitations on the total weight they can allocate across coalitions.

We first introduce formally the main notions used in this paper.
A \emph{set of agents} (or \emph{set of players}) is a nonempty finite set $I$.
For any set $I$ of agents, a \emph{set of coalitions} is a nonempty subset $E$ of the set $2^I \setminus \{\emptyset\}$ of nonempty subsets of $I$, and a \emph{budget vector} is a family $(b_i)_{i \in I}$ of positive real numbers.
For any set $I$ of agents, and any set $E$ of coalitions, a \emph{capacity vector} is a family $(c_e)_{e \in E}$ of real numbers in $[0,1]$.

\begin{definition}[Constrained network game]
    A \emph{constrained network game} is a tuple
    $$G = (I,E,(b_i)_{i \in I},(c_e)_{e \in E},(u_i)_{i \in I}),$$
    where $I$ is a set of agents, $E$ is a set of coalitions, $(b_i)_{i \in I}$ is a budget vector, $(c_e)_{e \in E}$ is a capacity vector, and, for any $i \in I$, $u_i$ is a function from
    $$X(G) = \prod_{e \in E} [0,c_e]$$
    to $\mathbb{R}$ called the \emph{payoff function of agent $i$}, where $X(G)$ is called the \emph{set of networks of $G$}. $\Diamond$
\end{definition}

Before continuing with our preliminaries, we introduce some practical notation that will be used throughout this paper.
Let $G$ be a constrained network game, and $e \in E$ be a coalition.
Define
$$E_{-e} = E \setminus \{e\} \quad \text{and} \quad X(G)_{-e} = \prod_{f \in E_{-e}} [0,c_f].$$
For any $d_e \in [0,c_e]$, and any $x_{-e} \in X(G)_{-e}$, $(d_e,x_{-e})$ is the network defined by $(d_e,x_{-e})_f = (x_{-e})_f$, for any $f \in E_{-e}$, and $(d_e,x_{-e})_e = d_e$.
For any network $x \in X(G)$, $x_{-e} \in X(G)_{-e}$ corresponds to the vector of weights obtained from $x$ by removing the coordinate associated with $e$.

Finally, let us introduce some properties for constrained network games that will be used for some existence and non-existence results.
A constrained network game $G$ is called:
\begin{enumerate}
    \item \emph{continuous} (resp. \emph{linear}) if, for any $i \in I$, $u_i$ is continuous (resp. linear).
    \item \emph{componentwise monotonic} (resp. \emph{componentwise quasiconcave}, resp. \emph{componentwise strictly monotonic}) if, for any $e \in E$, any $x_{-e} \in X(G)_{-e}$, and any $i \in e$, the function $u_i(\cdot,x_{-e}): x_e \mapsto u_i(x_e,x_{-e})$ is monotonic (resp. quasiconcave, resp. strictly monotonic).
    \item a \emph{network game} if:
    \begin{enumerate}
        \item $E = \{e \in 2^I \setminus \{\emptyset\} : \lvert e \rvert = 2\}$ (each coalition $e = \{i,j\} \in E$ is simply denoted $ij$).
        \item For any $i \in I$, $b_i = +\infty$.
        \item For any $e \in E$, $c_e = 1$.
    \end{enumerate}
\end{enumerate}

\section{Coalitionwise Stability}

Within the current framework, we now introduce a notion of stability -- called \emph{coalitionwise stability} -- that naturally extends pairwise stability to constrained network games.
The core intuition of pairwise stability is preserved: a unilateral decrease in a coalition's weight is feasible whenever at least one member benefits from it, whereas a joint increase can occur only if all members agree to it and the resulting allocation remains compatible with their individual constraints.
This approach allows us to explicitly account for constrained agents in the analysis of network formation.

We first recall the notion of \emph{pairwise stability} introduced by Bich and Morhaim \cite{bich_morhaim_2020} for network games. This concept extends the classical notion of pairwise stability of Jackson and Wolinsky \cite{jackson_wolinsky_1996} to environments in which links may take arbitrary intensities in the interval $[0,1]$.

\begin{definition*}[Bich--Morhaim \cite{bich_morhaim_2020}]
    Let $G$ be a network game.
    A network $x \in X(G)$ is \emph{pairwise stable with respect to $G$} if, for every $ij \in E$, the following conditions hold:
    \begin{enumerate}
        \item (Unilateral decrease)\textbf{.} For every $d_{ij} \in [0,x_{ij})$,
        $$u_i(d_{ij},x_{-ij}) \leq u_i(x) \quad \text{and} \quad u_j(d_{ij},x_{-ij}) \leq u_j(x).$$
        \item (Bilateral increase)\textbf{.} For every $d_{ij} \in (x_{ij},1]$,
        $$u_i(d_{ij},x_{-ij}) \leq u_i(x) \quad \text{or} \quad u_j(d_{ij},x_{-ij}) \leq u_j(x). \quad \Diamond$$
    \end{enumerate}
\end{definition*}

The main contribution of Bich and Morhaim \cite{bich_morhaim_2020} is to establish the existence of pairwise stable networks under standard continuity and componentwise quasiconcavity assumptions. Their result provides a natural benchmark for the constrained framework studied in this paper.

\begin{theorem*}[Bich--Morhaim \cite{bich_morhaim_2020}]
    Any continuous and componentwise quasiconcave network game admits a pairwise stable network.
\end{theorem*}

When budget constraints are introduced, however, the notion of pairwise stability must be adapted. Indeed, increasing the intensity of a given coalition may become impossible because of the resources already allocated to other coalitions involving the same agents.
Consequently, feasible deviations depend not only on the coalition under consideration, but also on the current network configuration.
To account for these additional feasibility restrictions, we introduce the following notion of stability.

\begin{definition}[Set of feasible networks]
    Let $G$ be a constrained network game.
    The set
    $$\mathcal{F}(G) = \left\{x \in X(G) : \forall i \in I, \sum_{e \ni i} x_e \leq b_i\right\}$$
    is called the \emph{set of feasible networks of $G$}. $\Diamond$
\end{definition}

\begin{definition}[Coalitionwise stable network]
    Let $G$ be a constrained network game.
    A feasible network $x \in \mathcal{F}(G)$ is \emph{coalitionwise stable with respect to $G$} if, for every $e \in E$, the following conditions hold:
    \begin{enumerate}
        \item (Unilateral decrease)\textbf{.} For every $d_e \in [0,x_e)$, and every $i \in e$,
        $$u_i(d_e,x_{-e}) \leq u_i(x).$$
        \item (Multilateral increase)\textbf{.} For every $d_e \in (x_e,c_e]$, if $(d_e,x_{-e}) \in \mathcal{F}(G)$, then there exists $i \in e$ such that
        $$u_i(d_e,x_{-e}) \leq u_i(x).$$
    \end{enumerate}
    The set of coalitionwise stable networks of $G$ is denoted by $\mathcal{E}(G)$. $\Diamond$
\end{definition}

Our first result shows that the existence theorem of Bich and Morhaim extends to the constrained setting. Despite the presence of budget restrictions and endogenous feasibility constraints, coalitionwise stable networks continue to exist under the same continuity and componentwise quasiconcavity assumptions.

\begin{theorem} \label{thm_1}
    Any continuous and componentwise quasiconcave constrained network game admits an coalitionwise stable network.
\end{theorem}

\begin{proof}
	See \ref{pf_thm_1}.
\end{proof}

Finally, we establish a useful characterization of coalitionwise stability, which will play a central role in the remainder of this paper. The proof is straightforward and therefore omitted.

\begin{proposition}
    Let $G$ be a constrained network game, and $x \in \mathcal{F}(G)$ be a feasible network.
    Then, $x \in \mathcal{E}(G)$ if and only if, for every $e \in E$, the following conditions hold:
    \begin{enumerate}
        \item For every $d_e \in [0,x_e)$, and every $i \in e$,
        $$u_i(d_e,x_{-e}) \leq u_i(x).$$
        \item For every $d_e \in (x_e,\mu_e(G,x_{-e})]$, there exists $i \in e$ such that
        $$u_i(d_e,x_{-e}) \leq u_i(x),$$
        where
        $$\mu_e(G,x_{-e}) = \min \left\{c_e,\min_{j \in e} \left(b_j - \sum_{\substack{f \ni j\\f \neq e}} x_f\right)\right\}.$$
    \end{enumerate}
\end{proposition}

The previous proposition reformulates coalitionwise stability in a way that is better suited for perturbation arguments. In particular, considering a constrained network game $G$, $e \in E$, and $x_{-e} \in X(G)_{-e}$, the quantity $\mu_e(G,x_{-e})$ describes the maximal feasible increase of coalition $e$ given the actions chosen by the other coalitions. This characterization will serve as the benchmark notion of stability throughout the next section.

\section{Perfect Equilibria in Network Formation Theory}

The notion of perfection is due to Selten \cite{selten_1975} and his work on equilibrium refinements in finite normal-form games.
In order to eliminate non-robust Nash equilibria sustained by weakly dominated behaviors, Selten introduced the notion of \emph{perfect equilibrium} by perturbing the strategic environment and requiring every pure strategy to remain played with a small positive probability.

Recall that a \emph{(finite) normal-form game} is a tuple
$$\Gamma = (I,(S_i)_{i \in I},(\pi_i)_{i \in I})$$
where $I$ is a set of players, and, for any $i \in I$, $S_i$ is a nonempty (finite) set called the \emph{strategy set of player $i$} and $\pi_i$ is a function from $S = \prod_{i \in I} S_i$ to $\mathbb{R}$ called the \emph{payoff function of player $i$}.

Also, given a finite normal-form game $\Gamma$, recall that the \emph{mixed extension of $\Gamma$} is the normal-form game
$$(I,(\Sigma_i)_{i \in I},(\Pi_i)_{i \in I}),$$
where, for every $i \in I$, $\Sigma_i$ is the set of probability measures on $S_i$ and $\Pi_i$ is the function from $\Sigma = \prod_{i \in I} \Sigma_i$ to $\mathbb{R}$ defined as follows: for every $\sigma \in \Sigma$,
$$\Pi_i(\sigma) = \sum_{s \in S} \pi_i(s) \prod_{j \in I} \sigma_j(s_j).$$

\begin{definition*}[Selten \cite{selten_1975}]
    Let $\Gamma$ be a finite normal-form game.
    A strategy profile $\sigma \in \Sigma$ is a \emph{perfect equilibrium of $\Gamma$} if there exists a sequence $(\eta^n,\sigma^n)$ in $\prod_{i \in I} \mathbb{R}^{S_i} \times \Sigma$ that converges to $(\mathbf{0},\sigma)$ such that, for every $n$:
    \begin{enumerate}
        \item For every $i \in I$, $\sum_{s_i \in S_i} \eta^n_i(s_i) < 1$ and $\eta^n_i(s_i) > 0$, for all $s_i \in S_i$.
        \item For every $i \in I$, $\sigma^n_i$ belongs to
        $$\Sigma^n_i = \{\tau_i \in \Sigma_i : \forall s_i \in S_i, \tau_i(s_i) \geq \eta^n_i(s_i)\}.$$
        \item $\sigma^n$ is a Nash equilibrium of $(I,(\Sigma^n_i)_{i \in I},(\Pi^n_i)_{i \in I})$, where, for any $i \in I$, $\Pi^n_i$ is the restriction of $\Pi_i$ to $\prod_{i \in I} \Sigma^n_i$. $\Diamond$
    \end{enumerate}
\end{definition*}

\begin{theorem*}[Selten \cite{selten_1975}]
    Any finite normal-form game admits a perfect equilibrium.
\end{theorem*}

As also recalled in the introduction, Selten \cite{selten_1975} subsequently proved that this perturbative definition is equivalent to a more behavioral characterization, where players optimize against nearby completely mixed strategy profiles.

\begin{theorem*}[Selten \cite{selten_1975}]
    Let $\Gamma$ be a finite normal-form game.
    A strategy profile $\sigma \in \Sigma$ is a perfect equilibrium of $\Gamma$ if and only if there exists a sequence $(\sigma^n)$ in $\Sigma$ that converges to $\sigma$ such that, for every $n$, and every $i \in I$, the following conditions hold:
    \begin{enumerate}
        \item $\sigma^n$ belongs to $\mathrm{int}(\Sigma)$.
        \item For every $\tau_i \in \Sigma_i$,
        $$\Pi_i(\tau_i,\sigma^n_{-i}) \leq \Pi_i(\sigma_i,\sigma^n_{-i}).$$
    \end{enumerate}
\end{theorem*}

Building upon this idea, Bich and Teteryatnikova \cite{bich_teteryatnikova_2023} transposed Selten's approach of perfection to the framework of network formation theory.
They introduced the notion of \emph{perfect pairwise stability} (here called \emph{trembling-hand equilibrium}), which refines pairwise stability by requiring every link decision to remain stable under small perturbations of the other link weights.

\begin{definition*}[Bich--Teteryatnikova \cite{bich_teteryatnikova_2023}]
    Let $G$ be a network game.
    A network $x \in X(G)$ is a \emph{trembling-hand equilibrium of $G$} if and only if there exists a sequence $(x^n)$ in $X(G)$ that converges to $x$ such that, for every $n$, and every $ij \in E$, the following conditions hold:
    \begin{enumerate}
        \item $x^n$ belongs to $\mathrm{int}(X(G))$.
        \item For every $d_{ij} \in [0,x_{ij})$, $$u_i(d_{ij},x^n_{-ij}) \leq u_i(x_{ij},x^n_{-ij}) \quad \text{and} \quad u_j(d_{ij},x^n_{-ij}) \leq u_j(x_{ij},x^n_{-ij}).$$
        \item For every $d_{ij} \in (x_{ij},1]$, $$u_i(d_{ij},x^n_{-ij}) \leq u_i(x_{ij},x^n_{-ij}) \quad \text{or} \quad u_j(d_{ij},x^n_{-ij}) \leq u_j(x_{ij},x^n_{-ij}). \quad \Diamond$$
    \end{enumerate}
\end{definition*}

In the same paper, Bich and Teteryatnikova \cite{bich_teteryatnikova_2023} established an existence theorem for this notion, under an additional assumption of monotonicity.

\begin{theorem*}[Bich--Teteryatnikova \cite{bich_teteryatnikova_2023}]
    Any continuous, componentwise quasiconcave and componentwise strictly monotonic network game admits a trembling-hand equilibrium.
\end{theorem*}

They also provided, for componentwise strictly monotonic network games, an equivalence between this trembling-hand formulation and a perturbative characterization involving pairwise stable networks in suitably perturbed environments.

\begin{definition*}[Bich--Teteryatnikova \cite{bich_teteryatnikova_2023}]
    Let $G$ be a network game.
    A network $x \in X(G)$ is \emph{a perturbed equilibrium of $G$} if there exists a sequence $(\varepsilon^n,x^n)$ in $X(G) \times X(G)$ that converges to $(\mathbf{0},x)$ such that, for every $n$, and every $ij \in E$, the following conditions hold:
    \begin{enumerate}
        \item $0 < \varepsilon^n_{ij} < \frac{1}{2}$.
        \item $x^n_{ij}$ belongs to $[\varepsilon^n_{ij},1-\varepsilon^n_{ij}]$.
        \item For every $d^n_{ij} \in [\varepsilon^n_{ij},x^n_{ij})$,
        $$u_i(d^n_{ij},x^n_{-ij}) \leq u_i(x^n) \quad \text{and} \quad u_j(d^n_{ij},x^n_{-ij}) \leq u_j(x^n).$$
        \item For every $d^n_{ij} \in (x^n_{ij},1-\varepsilon^n_{ij}]$,
        $$u_i(d^n_{ij},x^n_{-ij}) \leq u_i(x^n) \quad \text{or} \quad u_j(d^n_{ij},x^n_{-ij}) \leq u_j(x^n). \quad \Diamond$$
    \end{enumerate}
\end{definition*}

\begin{theorem*}[Bich--Teteryatnikova \cite{bich_teteryatnikova_2023}]
    Let $G$ be a componentwise strictly monotonic network game.
    A network is a trembling-hand equilibrium of $G$ if and only if it is a perturbed equilibrium of $G$.\footnote{Bich and Teteryatnikova state their original existence theorem under continuity, componentwise quasiconcavity, and componentwise monotonicity (see Theorem 3 in \cite{bich_teteryatnikova_2023}). \emph{However, one step of their proof appears to require in fact a stronger strict monotonicity assumption.} We therefore establish below a constrained analogue under componentwise strict monotonicity (see Theorem \ref{thm_4}).}
\end{theorem*}

The next two subsections are devoted to extending these ideas to constrained network games.
In this setting, the feasible set no longer has a product structure, which creates substantial additional difficulties for trembling-hand type refinements. Our objective is therefore to study constrained analogues of both notions introduced by Bich and Teteryatnikova \cite{bich_teteryatnikova_2023}: a perturbative notion and a trembling-hand notion of perfect stability.

\emph{Throughout the remainder of this paper, we suppose that, when we consider a constrained network game $G$, the set $\mathcal{F}(G)$ has a nonempty interior.}

\subsection{Perturbed Equilibria Under Constraints}

We start by introducing the perturbative formulation of perfection.
While the trembling-hand approach introduced in the next subsection perturbs the surrounding network and evaluates the robustness of a fixed limit action, the perturbative approach modifies the feasible environment itself. This is closer in spirit to Selten's original definition of perfect equilibrium, where equilibria are obtained as limits of equilibria of perturbed games.

In the constrained setting, perturbing the environment requires modifying both types of feasibility restrictions. First, each coalition weight is bounded away from zero and from its maximal capacity. Second, each agent is required to keep a small amount of budget unallocated. The resulting feasible set is therefore a strict inner approximation of the original feasible set.

\begin{definition}[Perturbed equilibrium]
    Let $G$ be a constrained network game.
    A feasible network $x \in \mathcal{F}(G)$ is a \emph{perturbed equilibrium of $G$} if there exists a sequence $(\varepsilon^n,\eta^n,x^n)$ in $X(G) \times [0,+\infty)^I \times X(G)$ that converges to $(\mathbf{0},\mathbf{0},x)$ such that, for every $n$, the following conditions hold:
    \begin{enumerate}
        \item For every $i \in I$, $0 < \eta^n_i < b_i$.
        \item For every $e \in E$, $0 < \varepsilon^n_e < \frac{c_e}{2}$.
        \item $x^n$ belongs to
        $$\mathcal{F}^n(G) = \left\{z^n \in \prod_{e \in E} [\varepsilon^n_e,c_e-\varepsilon^n_e] : \forall i \in I, \sum_{e \ni i} z^n_e \leq b_i - \eta^n_i\right\}.$$
        \item For every $e \in E$:
        \begin{enumerate}
            \item (Unilateral decrease)\textbf{.} For every $d^n_e \in [\varepsilon^n_e,x^n_e)$, and every $i \in e$,
            $$u_i(d^n_e,x^n_{-e}) \leq u_i(x^n).$$
            \item (Multilateral increase)\textbf{.} For every $d^n_e \in (x^n_e,\mu^n_e(G,x^n_{-e})]$, there exists $i \in e$ such that
            $$u_i(d^n_e,x^n_{-e}) \leq u_i(x^n),$$
            where
            $$\mu^n_e(G,x^n_{-e}) = \min \left\{c_e-\varepsilon^n_e,\min_{j \in e} \left(b_j - \eta^n_j - \sum_{\substack{f \ni j\\f \neq e}} x^n_f\right)\right\}.$$
        \end{enumerate}
    \end{enumerate}
    The set of perturbed equilibria of $G$ is denoted by $\mathcal{P}(G)$. $\Diamond$
\end{definition}

Thus, a perturbed equilibrium is a limit of networks that are stable in slightly restricted feasible environments. The capacity perturbation parameters prevent coalition weights from reaching the boundary of their capacity constraints, while the budget parameters keep agents away from exhausting their budgets. As these perturbations vanish, the corresponding stable networks converge to a network of the original game.

\begin{figure}[htbp]
    \centering
    \begin{tikzpicture}[
        scale=5,
        >=Latex,
        every node/.style={font=\small}
    ]

        \def\epsE{0.25}
        \def\epsF{0.25}
        \def\eta{0.05}

        \fill[gray!15]
            (0,0) -- (1,0) -- (0,1) -- cycle;

        \coordinate (A) at (\epsE,\epsF);
        \coordinate (B) at ({1-\eta-\epsF},\epsF);
        \coordinate (C) at (\epsE,{1-\eta-\epsE});

        \fill[gray!38]
            (A) -- (B) -- (C) -- cycle;

        \draw[thick] (0,0) rectangle (1,1);
        \draw[thick] (0,1) -- (1,0);

        \draw[thick]
            (A) -- (B) -- (C) -- cycle;

        \draw[->] (-0.03,0) -- (1.12,0)
            node[right] {$x_e$};
        \draw[->] (0,-0.03) -- (0,1.12)
            node[above] {$x_f$};

        \node[below left] at (0,0) {$0$};
        \node[below] at (1,0) {$c_e$};
        \node[left] at (0,1) {$c_f$};

        \node at (0.20,0.15) {$\mathcal{F}(G)$};
        \node at (0.40,0.35) {$\mathcal{F}^n(G)$};
        \node at (0.80,0.75) {$X(G)$};

    \end{tikzpicture}

    \caption{
        Geometry of the perturbative approach.
        The feasible set $\mathcal{F}(G)$ does not possess a product structure.
        The perturbed feasible sets $\mathcal{F}^{n}(G)$ are strict
        inner approximations obtained by imposing positive lower bounds
        on coalition intensities and requiring agents to leave a positive
        amount of their budgets unallocated.
    }
    \label{fig_1}
\end{figure}
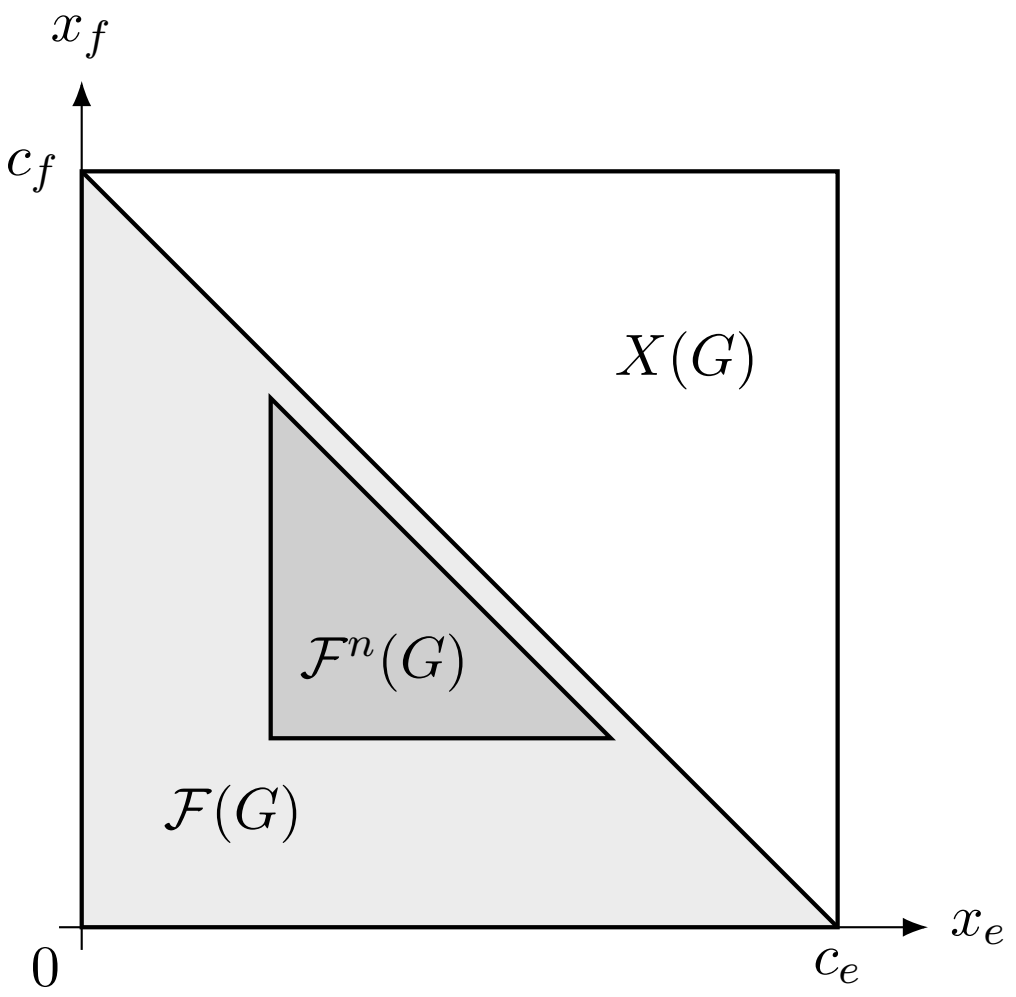

Figure \ref{fig_1} illustrates the geometry of the perturbative approach.

The next proposition shows that perturbed perfection is a refinement of coalitionwise stability.

\begin{proposition}
    Let $G$ be a continuous constrained network game.
    If a network is a perturbed equilibrium of $G$, then it is coalitionwise stable with respect to $G$.
\end{proposition}

\begin{proof}
    Let $x \in \mathcal{P}(G)$, and consider a corresponding perturbing sequence $(\varepsilon^n,\eta^n,x^n)$.
    Fix a coalition $e \in E$.
    We first check the unilateral decrease condition.
    By contradiction, suppose that there exists $d_e \in [0,x_e)$ and $i \in e$ such that $u_i(d_e,x_{-e}) > u_i(x)$.
    Suppose first that $d_e > 0$.
    Since $u_i$ is continuous and $x^n \longrightarrow x$, there exists $\bar n_1$ such that, for every $n \geq \bar n_1$, $u_i(d_e,x^n_{-e}) > u_i(x^n)$.
    Since $d_e < x_e$ and $x^n \longrightarrow x$, there exists $\bar n_2$ such that, for every $n \geq \bar n_2$, $d_e < x^n_e$.
    Since $0 < d_e$ and $\varepsilon^n \longrightarrow \mathbf{0}$, there exists $\bar n_3$ such that, for every $n \geq \bar n_3$, $\varepsilon^n_e < d_e$.
    Hence, we obtain a contradiction with the unilateral decrease condition in the definition of a perturbed equilibrium, for $\bar n = \max \{\bar n_1,\bar n_2,\bar n_3\}$.
    Suppose now that $d_e = 0$.
    Since $u_i$ is continuous, $x^n \longrightarrow x$ and $\varepsilon^n \longrightarrow \mathbf{0}$, there exists $\bar n$ such that, for every $n \geq \bar n$, $u_i(\varepsilon^n_e,x^n_{-e}) > u_i(x^n)$, which directly leads to a contradiction with the unilateral decrease condition in the definition of a perturbed equilibrium.
    In order to check the multilateral increase condition, observe that the reasoning is similar, noticing that $\mu^n_e(G,x^n_{-e}) \longrightarrow \mu_e(G,x_{-e})$.
\end{proof}

The following theorem establishes the existence of perturbed equilibria under the usual assumptions.

\begin{theorem} \label{thm_3}
    Any continuous and componentwise quasiconcave constrained network game admits a perturbed equilibrium.
\end{theorem}

\begin{proof}
	See \ref{pf_thm_3}.
\end{proof}

In the next subsection, we will see that under some strict monotonicity assumption, the perturbative notion is equivalent to a certain kind of trembling-hand formulation (see Theorem \ref{thm_4}).
However, we can still provide an example of a continuous and componentwise quasiconcave constrained network game that is not componentwise strictly monotonic, and such that its set of perturbed equilibria is strictly included in its set of coalitionwise stable networks.

\begin{example}
    Consider the constrained network game $G$, where
    $I = \{1,2,3\}$, $E = \{e,f\}$, with $e = \{1,2\}$ and
    $f = \{1,3\}$, $b_1 = b_2 = b_3 = 1$, and $c_e = c_f = 1$, and
    $$u_1(x) = 0, \quad u_2(x)=- x_ex_f, \quad u_3(x)=x_f, \quad \forall x = (x_e,x_f) \in X(G).$$
    Clearly, $G$ is continuous and componentwise quasiconcave, but not componentwise strictly monotonic.
    We claim that
    $$\mathcal{E}(G) = \{(x_e,x_f) \in \mathcal{F}(G) : x_e = 0 \text{ or } x_f = 0\}.$$
    Fix $x \in \mathcal{F}(G)$.
    If $x_e > 0$ and $x_f > 0$, then agent 2 strictly benefits from decreasing $x_e$.
    Thus such a network cannot be coalitionwise stable.
    Conversely, suppose that $x_e = 0$ or $x_f = 0$.
    Consider first coalition $e$: no profitable unilateral decrease is possible (payoffs of agents 1 and 2 are both equal to $0$), while increases are blocked by agent 1 (whose payoff is constant).
    Consider now coalition $f$: decreases are never profitable for agent 3, and increases are again blocked by agent 1.
    We now show that
    $$\mathcal{P}(G) = \{(0,x_f) : x_f \in [0,1]\} \subset \mathcal{E}(G).$$
    Fix $x \in \mathcal{F}(G)$.
    Suppose first that $x$ is a perturbed equilibrium of $G$, and let
    $(\varepsilon^n,\eta^n,x^n)$ be a corresponding perturbing sequence.
    By definition, for any $n$, we have that $x_f^n \geq \varepsilon_f^n > 0$.
    Hence, necessarily $x_e^n = \varepsilon_e^n$, for all $n$ (otherwise, agent 2 would have a strict incentive to decrease $x^n_e$, which would lead to a contradiction with $x$ being a perturbed equilibrium of $G$).
    Since $\varepsilon^n \longrightarrow \mathbf{0}$, we obtain that $x = \lim_n (x^n_e,x^n_f) = \lim_n (\varepsilon^n_e,x^n_f) = (0,x_f)$.
    Conversely, suppose that $x_e = 0$.
    Consider a sequence $(\delta^n)$ in $[0,\frac{1}{3}]$ that converges to $0$, and such that $0 < \delta^n < \frac{1}{3}$, for all $n$.
    Now, for any $n$, define $\varepsilon^n_e = \varepsilon^n_f = \eta^n_1 = \eta^n_2 = \eta^n_3 = x^n_e = \delta^n$ and $x_f^n = (1-3\delta^n)x_f + \delta^n$.
    A direct verification shows that the sequence $(\varepsilon^n,\eta^n,x^n)$ satisfies the required properties. $\Diamond$
\end{example}

\subsection{Trembling-Hand Equilibria Under Constraints}

In unconstrained environments, trembling-hand perfection captures the idea that a stable network should remain stable against small perturbations of the surrounding links. When budget constraints are present, however, the feasible set depends on the actions chosen by other agents, which creates additional difficulties. In particular, perturbing a network may modify the set of feasible deviations available to a coalition.

We begin by introducing a first notion of \emph{feasible trembling-hand equilibrium} that closely follows the original definition. We then show that this notion may fail to exist even in simple linear constrained network games. This motivates a weaker definition, called \emph{semi-feasible trembling-hand equilibrium}, which relaxes some feasibility requirements. We prove that existence may still fail. Finally, we introduce a third notion of \emph{trembling-hand equilibrium} and establish an existence theorem under standard continuity, monotonicity, and quasiconcavity assumptions.

A first natural extension of trembling-hand perfection consists in requiring that, after small perturbations of the surrounding network, the reference action remains feasible and continues to satisfy the stability conditions. This leads to the following definition.

\begin{definition}[Feasible trembling-hand equilibrium]
    Let $G$ be a constrained network game.
    A feasible network $x \in \mathcal{F}(G)$ is a \emph{feasible trembling-hand equilibrium of $G$} if there exists a sequence $(x^n)$ in $\mathcal{F}(G)$ that converges to $x$ such that, for every $n$, and every $e \in E$, the following conditions hold:
    \begin{enumerate}
        \item $x^n$ belongs to $\mathrm{int}(\mathcal{F}(G))$.
        \item $(x_e,x^n_{-e})$ belongs to $\mathcal{F}(G)$.
        \item For every $d_e \in [0,x_e)$, and every $i \in e$,
        $$u_i(d_e,x^n_{-e}) \leq u_i(x_e,x^n_{-e}).$$
        \item For every $d_e \in (x_e,\mu_e(G,x^n_{-e})]$, there exists $i \in e$ such that
        $$u_i(d_e,x^n_{-e}) \leq u_i(x_e,x^n_{-e}),$$
        where
        $$\mu_e(G,x^n_{-e}) = \min \left\{c_e,\min_{j \in e} \left(b_j - \sum_{\substack{f \ni j\\f \neq e}} x^n_f\right)\right\}. \quad \Diamond$$
    \end{enumerate}
\end{definition}

The next result shows that feasible trembling-hand perfection is indeed a refinement of coalitionwise stability.

\begin{proposition} \label{prop_3}
    Let $G$ be a continuous constrained network game.
    If a network is a feasible trembling-hand equilibrium of $G$, then it is coalitionwise stable with respect to $G$.
\end{proposition}

\begin{proof}
    Let $\bar x \in \mathcal{F}(G)$ be a feasible trembling-hand equilibrium, and let $(x^n)$ be a sequence in $\mathcal{F}(G)$ as in the definition.
    Fix $e \in E$. We consider three cases.
    First, let $d_e \in [0,\bar x_e)$, and $i \in e$.
    For every $n$, we have that $u_i(d_e,x^n_{-e}) \leq u_i(\bar x_e,x^n_{-e})$, by definition of $\bar x$.
    Since $x^n \longrightarrow \bar x$, and since $u_i$ is continuous, passing to the limit directly gives $u_i(d_e,\bar x_{-e}) \leq u_i(\bar x)$.
    Second, let $d_e \in (\bar x_e,\mu_e(G,\bar x_{-e}))$.
    By continuity of $x_{-e} \mapsto \mu_e(G,x_{-e})$, we have that there exists $\bar n$ such that, for every $n \geq \bar n$, $d_e < \mu_e(G,x^n_{-e})$.
    Because $\bar x$ is a feasible trembling-hand equilibrium, for every $n \geq \bar n$, there exists $i_n \in e$ such that $u_{i_n}(d_e,x^n_{-e}) \leq u_{i_n}(\bar x_e,x^n_{-e})$.
    Since $e$ is finite, consider a subsequence $(x^{n_m})$ starting after $\bar n$ and such that $m \mapsto i_{n_m}$ is constantly equal to some $i \in e$.
    Using continuity of $u_i$ and passing to the limit yields $u_i(d_e,\bar x_{-e}) \leq u_i(\bar x)$.
    Finally, consider a sequence $(d^p_e)$ in $(\bar x_e,\mu_e(G,\bar x_{-e})]$ such that $d^p_e < \mu_e(G,\bar x_{-e})$, for all $p$, and that converges to $\mu_e(G,\bar x_{-e})$.
    Using the result of the previous case, for any $p$, there exists $i_p \in e$ such that $u_{i_p}(d^p_e,\bar x_{-e}) \leq u_{i_p}(\bar x)$.
    Since $e$ is finite, we can use the previous argument again in order to obtain that $u_i(\mu_e(G,\bar x_{-e}),\bar x_{-e}) \leq u_i(\bar x)$.
\end{proof}

Unfortunately, the previous notion turns out to be too demanding in constrained environments. The reason is that perturbations of the surrounding links may destroy the feasibility of the reference action itself. The following example illustrates this phenomenon.

\begin{proposition} \label{prop_1}
    There exists a linear constrained network game that admits no feasible trembling-hand equilibrium.
\end{proposition}

\begin{proof}
    Consider the network game $G$, where $I = \{1,2,3\}$, $E = \{e,f\}$, with $e = \{1,2\}$ and $f = \{1,3\}$, $b_1 = b_2 = b_3 = 1$, $c_e = c_f = 1$, and
    $$u_1(x) = x_e-x_f, \quad u_2(x) = x_e, \quad u_3(x) = -x_f, \quad \forall x = (x_e,x_f) \in X(G).$$
    The only coalitionwise stable network here is $\bar x = (1,0)$.
    However, $\bar x$ is not a feasible trembling-hand equilibrium.
    Indeed, if there existed a sequence $(x^n)$ in $\mathcal{F}(G)$ as in the definition, then we should have in particular that, for any $n$, $(\bar x_e,x^n_{-e})$ belongs to $\mathcal{F}(G)$, i.e., that $1 + x^n_f \leq 1$, implying that $x^n_{f} = 0$.
    This leads to a contradiction with $x^n \in \mathrm{int}(\mathcal{F}(G))$, for all $n$.
\end{proof}

\begin{figure}[htbp]
    \centering
    \begin{tikzpicture}[
        scale=5,
        >=Latex,
        every node/.style={font=\small}
    ]

        \fill[gray!15]
            (0,0) -- (1,0) -- (0,1) -- cycle;

        \draw[thick] (0,0) rectangle (1,1);
        \draw[thick] (0,1) -- (1,0);

        \draw[->] (-0.03,0) -- (1.15,0)
            node[right] {$x_e$};
        \draw[->] (0,-0.03) -- (0,1.12)
            node[above] {$x_f$};

        \node[below left] at (0,0) {$0$};
        \node[below] at (1,0) {$1$};
        \node[left] at (0,1) {$1$};

        \node at (0.25,0.35) {$\mathcal{F}(G)$};
        \node at (0.80,0.75) {$X(G)$};

        \coordinate (xn) at (0.6,0.3);
        \coordinate (xbar) at (1,0);
        \coordinate (hybrid) at (1,0.3);

        \fill (xn) circle (0.9pt);
        \fill (xbar) circle (0.9pt);
        \draw[thick,fill=white] (hybrid) circle (1.1pt);

        \node[above left=1pt] at (xn) {$x^n$};
        \node[below right=1pt] at (xbar)
            {$\bar x$};
        \node[right=2pt] at (hybrid)
            {$(\bar x_e,x_f^n)$};

        \draw[
            ->,
            thick,
            dashed
        ]
            ($(xn)+(0.02,0)$)
            --
            ($(hybrid)+(-0.02,0)$);

    \end{tikzpicture}

    \caption{
        Failure of the feasible trembling-hand formulation in Proposition \ref{prop_1}.
        Although $x^n$ is an interior feasible network converging to
        $\bar x = (1,0)$, replacing its $e$-coordinate by the limiting
        intensity $\bar x_e = 1$ produces the infeasible network $(\bar x_e,x_f^n)$.
    }
    \label{fig_2}
\end{figure}
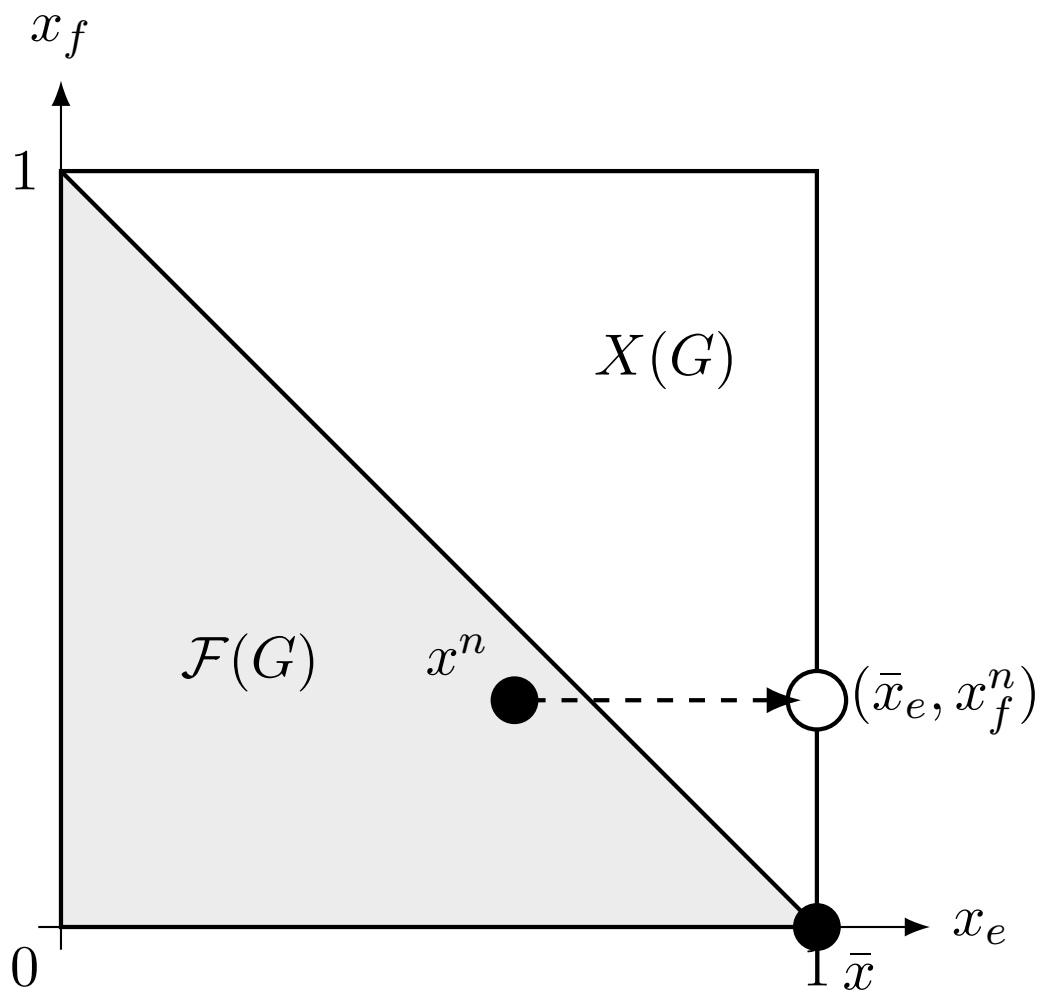

The difficulty underlying the feasible trembling-hand formulation identified in Proposition \ref{prop_1} is illustrated in Figure \ref{fig_2}.

To overcome this issue, one may relax the feasibility requirement and only keep the optimality conditions. This leads to the following weaker notion.

\begin{definition}[Semi-feasible trembling-hand equilibrium]
    Let $G$ be a constrained network game.
    A feasible network $x \in \mathcal{F}(G)$ is a \emph{semi-feasible trembling-hand equilibrium of $G$} if there exists a sequence $(x^n)$ in $\mathcal{F}(G)$ that converges to $x$ such that, for every $n$, and every $e \in E$, the following conditions hold:
    \begin{enumerate}
        \item $x^n$ belongs to $\mathrm{int}(\mathcal{F}(G))$.
        \item For every $d_e \in [0,x_e)$, and every $i \in e$,
        $$u_i(d_e,x^n_{-e}) \leq u_i(x_e,x^n_{-e}).$$
        \item For every $d_e \in (x_e,\mu_e(G,x^n_{-e})]$, there exists $i \in e$ such that
        $$u_i(d_e,x^n_{-e}) \leq u_i(x_e,x^n_{-e}),$$
        where
        $$\mu_e(G,x^n_{-e}) = \min \left\{c_e,\min_{j \in e} \left(b_j - \sum_{\substack{f \ni j\\f \neq e}} x^n_f\right)\right\}. \quad \Diamond$$
    \end{enumerate}
\end{definition}

As before, semi-feasible trembling-hand perfection refines coalitionwise stability (the proof is identical to the one of Proposition \ref{prop_3} and is therefore omitted).

\begin{proposition}
    Let $G$ be a continuous constrained network game.
    If a network is a semi-feasible trembling-hand equilibrium of $G$, then it is coalitionwise stable with respect to $G$.
\end{proposition}

One may now wonder whether removing the feasibility requirement is sufficient to recover existence. The next example shows that this is not the case.

\begin{proposition} \label{prop_2}
    There exists a linear constrained network game that admits no semi-feasible trembling-hand equilibrium.
\end{proposition}

\begin{proof}
    Consider the network game $G$, where $I = \{1,2,3\}$, $E = \{e,f\}$, with $e = \{1,2\}$ and $f = \{1,3\}$, $b_1 = b_2 = b_3 = 1$, $c_e = c_f = 1$, and
    $$u_1(x) = x_e+x_f, \quad u_2(x) = x_e, \quad u_3(x) = x_f, \quad \forall x = (x_e,x_f) \in X(G).$$
    Suppose that $G$ admits a semi-feasible trembling-hand equilibrium $\bar x \in \mathcal{F}(G)$.
    Consider a sequence $(x^n)$ in $\mathcal{F}(G)$ as in the definition, i.e., such that $x^n_e + x^n_f < 1$.
    We have that $\mu_e(G,x^n_{-e}) = 1 - x^n_f$ and $\mu_f(G,x^n_{-f}) = 1 - x^n_e$.
    Since $\bar x$ is a semi-feasible trembling-hand equilibrium, and since every payoff function is strictly increasing in its incident variables, we must have that $\bar x_e \geq \mu_e(G,x^n_{-e})$ and $\bar x_f \geq \mu_f(G,x^n_{-f})$.
    Summing the two previous inequalities leads to $\bar x_e + \bar x_f \geq 2 - (x^n_e + x^n_f) > 1$, which contradicts $\bar x \in \mathcal{F}(G)$.
\end{proof}

The previous proposition highlights a second source of difficulty. Even when the reference action is allowed to become infeasible after perturbations, the perturbation itself modifies the maximal feasible deviations. As a consequence, new profitable deviations may appear and prevent the existence of semi-feasible trembling-hand equilibria.

The failure of the previous notions comes from the fact that the feasible deviation set varies with the perturbed network. To avoid this issue, we evaluate deviations with respect to the feasible deviation set associated with the limit network itself. This leads to the following notion of trembling-hand equilibrium.

\begin{definition}[Trembling-hand equilibrium]
    Let $G$ be a constrained network game.
    A feasible network $x \in \mathcal{F}(G)$ is a \emph{trembling-hand equilibrium of $G$} if there exists a sequence $(x^n)$ in $\mathcal{F}(G)$ that converges to $x$ such that, for every $n$, and every $e \in E$, the following conditions hold:
    \begin{enumerate}
        \item $x^n$ belongs to $\mathrm{int}(\mathcal{F}(G))$.
        \item (Unilateral decrease)\textbf{.} For every $d_e \in [0,x_e)$, and every $i \in e$,
        $$u_i(d_e,x^n_{-e}) \leq u_i(x_e,x^n_{-e}).$$
        \item (Multilateral increase)\textbf{.} For every $d_e \in (x_e,\mu_e(G,x_{-e})]$, there exists $i \in e$ such that
        $$u_i(d_e,x^n_{-e}) \leq u_i(x_e,x^n_{-e}),$$
        where
        $$\mu_e(G,x_{-e}) = \min \left\{c_e,\min_{j \in e} \left(b_j - \sum_{\substack{f \ni j\\f \neq e}} x_f\right)\right\}.$$
    \end{enumerate}
    The set of trembling-hand equilibria of $G$ is denoted by $\mathcal{T}(G)$. $\Diamond$
\end{definition}

In a similar way, trembling-hand perfection is a refinement of coalitionwise stability (the proof is similar to the one of Proposition \ref{prop_3} and is also omitted).

\begin{proposition}
    Let $G$ be a continuous constrained network game.
    If a network is a trembling-hand equilibrium of $G$, then it is coalitionwise stable with respect to $G$.
\end{proposition}

We first compare the trembling-hand formulation with the perturbative formulation introduced in the previous subsection.

\begin{theorem} \label{thm_4}
    Let $G$ be a continuous and componentwise strictly monotonic constrained network game.
    A network is a perturbed equilibrium of $G$ if and only if it is a trembling-hand equilibrium of $G$.
\end{theorem}

\begin{proof}
	See \ref{pf_thm_4}.
\end{proof}

This equivalence shows that, under the appropriate assumptions, the behavioral trembling-hand formulation and the perturbative formulation are two complementary descriptions of the same robustness requirement.
In particular, unlike the previous notions, the concept of trembling-hand equilibrium admits an existence result under some strict monotonicity assumption.

\begin{corollary} \label{coro_1}
    Any continuous and componentwise strictly monotonic constrained network game admits a trembling-hand equilibrium.
\end{corollary}

\begin{proof}
    Since componentwise (strict) monotonicity implies componentwise quasiconcavity, the result follows directly from Theorem \ref{thm_3}.
\end{proof}

The strict monotonicity assumption plays a crucial role in the proof of Theorem \ref{thm_4}. However, as we shall see below, the existence result can be extended beyond strict monotonicity.
To state this extension, we introduce a weaker monotonicity requirement. The following condition allows payoff functions to exhibit flat regions, while requiring that the direction of monotonicity remains unchanged across profiles. This assumption is weaker than componentwise strict monotonicity (see Example \ref{ex_1} below) and still turns out to be sufficient to guarantee the existence of trembling-hand equilibria.
A constrained network game $G$ is called \emph{fixed-sign componentwise monotonic} if, for any $e \in E$, and any $i \in e$, the function $u_i(\cdot,x_{-e}): x_e \mapsto u_i(x_e,x_{-e})$ is either nonincreasing, for every $x_{-e} \in X(G)_{-e}$, or nondecreasing, for every $x_{-e} \in X(G)_{-e}$.

\begin{theorem} \label{thm_5}
    Any continuous and fixed-sign componentwise monotonic constrained network game admits a trembling-hand equilibrium.
\end{theorem}

\begin{proof}
	See \ref{pf_thm_5}.
\end{proof}

Theorem \ref{thm_5} shows that the existence of trembling-hand equilibria does not rely exclusively on strict monotonicity. What is essential is rather the stability of the direction of incentives associated with each coalition intensity. The fixed-sign monotonicity assumption allows for indifference regions and flat payoff profiles, while still preventing the changes in local incentives that may obstruct the perturbation argument. As a consequence, trembling-hand equilibria exist in a broader class of constrained network games than those covered by Corollary \ref{coro_1}.

\begin{example} \label{ex_1}
    Consider the network game $G$, where $I = \{1,2,3\}$, $E = \{e,f\}$, with $e = \{1,2\}$ and $f = \{1,3\}$, $b_1 = b_2 = b_3 = 1$, $c_e = c_f = 1$, and
    $$u_1(x) = u_2(x) = \min \left\{x_e,\frac{1}{2}\right\} + x_f, \quad u_3(x) = x_f, \quad \forall x = (x_e,x_f) \in X(G).$$
    Observe that $G$ is continuous, but not componentwise strictly monotonic.
    However, it is fixed-sign componentwise monotonic (each payoff function is nondecreasing in the incident variables).
    Moreover, one can verify that
    $$\mathcal{E}(G) = \mathcal{P}(G) = \mathcal{T}(G) = \{(x_e,x_f) \in X(G) : x_e + x_f = 1\}. \quad \Diamond$$
\end{example}

\section{Some Applications}

This section illustrates the applicability of our equilibrium concepts
through several classical models of network formation and strategic
interaction.
The first application revisits the co-author model of Jackson and
Wolinsky \cite{jackson_wolinsky_1996} in a constrained weighted
coalitional framework and provides explicit sufficient conditions for
the existence of perturbed and trembling-hand equilibria, together with
illustrative numerical examples.
We then discuss the weighted two-way flow model of Bala and Goyal
\cite{bala_goyal_2000} and the public-good provision model of Bramoullé
and Kranton \cite{bramoulle_kranton_2007}, showing how our general
existence results naturally apply to models exhibiting very different
interaction structures.

\subsection{The Extended Co-author Model}

The \emph{co-author model} introduced by Jackson and Wolinsky \cite{jackson_wolinsky_1996} provides a classical application of network formation theory.
In their framework, links represent bilateral collaborations between researchers, and the value generated by a collaboration depends on the number of projects in which the participating researchers are involved.
While highly insightful, this formulation does not take into account important features of modern scientific collaboration.
First, many research projects involve more than two researchers and are therefore more naturally represented by coalitions rather than bilateral links.
Second, collaborations often differ in intensity: some projects require a substantial commitment of time and resources, while others involve only occasional interaction.
Finally, researchers face natural resource constraints, such as limited time, attention, funding, or supervision capacity, which restrict the overall amount of collaborative activity they can sustain.
The constrained weighted network framework developed in this paper allows these features to be incorporated explicitly.
Coalitions represent multi-author projects, weights capture the intensity of collaboration, coalition capacities model technological or organizational limitations on individual projects, and budget constraints represent the limited resources available to researchers. The following model provides a simple adaptation of the co-author framework to this richer environment.

Consider a nonempty finite set $I$ of researchers, a nonempty set
$E \subset 2^I \setminus \{\emptyset\}$ of potential research projects, a budget vector
$(b_i)_{i \in I}$ and a capacity vector $(c_e)_{e \in E}$.
For every project $e \in E$, the variable
$x_e \in [0,c_e]$ denotes the intensity of collaboration devoted to
project $e$.
Each researcher $i \in I$ faces the limited research capacity $b_i$,
so that feasible collaboration profiles satisfy
$$\sum_{e \ni i} x_e \leq b_i.$$
The utility of researcher $i \in I$ is defined by
$$u_i(x) = \sum_{e \ni i} x_e v_{i,e}(s_e(x)) - \psi_i(s_i(x)),$$
where:
\begin{itemize}
    \item $s_i(x) = \sum_{e \ni i} x_e$ denotes the total research effort allocated by researcher $i$;
    \item for any $e \in E$, $s_e(x) = (s_j(x))_{j \in e}$ is the workload profile of the members of project $e$;
    \item for any $e \ni i$, $v_{i,e}: \mathbb{R}^e_+ \to \mathbb{R}$ represents the scientific value that researcher $i$ derives from project $e$ as a function of the workloads of its participants;
    \item $\psi_i: \mathbb{R}_+ \to \mathbb{R}$ represents the cost incurred by researcher $i$ as a function of his total research effort.
\end{itemize}

We now study two specifications of our extension of the co-author model.

\subsubsection{The Linear Congestion Co-author Game}

\paragraph{Description of the game}

This first specification assumes that the scientific value generated by a research project decreases linearly with the overall workload of its participants.
It therefore captures situations in which each additional collaboration slightly reduces the productivity of all the other projects involving the same researchers.

Suppose that, for every $e \in E$, and every $i \in e$,
$$v_{i,e}: (s_j)_{j \in e} \longmapsto \alpha_{i,e}-\beta_{i,e} \sum_{j \in e} s_j,$$
where $\alpha_{i,e} \geq 0$ represents the intrinsic scientific value of project $e$ for researcher $i$, and $\beta_{i,e} \geq 0$ measures the sensitivity of this project to congestion.
Moreover, suppose that, for every $i \in I$,
$$\psi_i: s \longmapsto \gamma_i s + \kappa_i s^2,$$
where $\gamma_i \geq 0$ and $\kappa_i \geq 0$, reflects the fact that coordinating an increasing number of collaborations becomes progressively more demanding in terms of time, supervision, and cognitive effort.
The resulting utility for researcher $i \in I$ is therefore:
$$u_i(x) = \sum_{e \ni i} x_e \left(\alpha_{i,e}-\beta_{i,e} \sum_{j \in e} s_j(x)\right) - \gamma_i s_i(x) - \kappa_i s_i(x)^2.$$
We refer to this specification as the \emph{linear congestion co-author game}, and denote this constrained network game by $G^\mathrm{lc}$.

We first observe that this specification satisfies the assumptions of Theorem \ref{thm_3}.

\begin{proposition} \label{prop_co-auth_1}
    The linear congestion co-author game is continuous and componentwise quasiconcave.
    Consequently, it admits a perturbed equilibrium, and therefore an coalitionwise stable network.
\end{proposition}

\begin{proof}
    See \ref{pf_prop_co-auth_1}.
\end{proof}

We can also give an explicit sufficient condition under which the same specification satisfies the monotonicity assumption used in Theorem \ref{thm_5}.
For every $i \in I$, define
$$\mathfrak{m}_i = \lvert \{e \in E : i \in e\} \rvert \quad \text{and} \quad \hat \beta_i = \max_{e \ni i} \beta_{i,e},$$
and define
$$\mathfrak{m} = \max_{j \in I} \mathfrak{m}_j.$$

\begin{proposition} \label{prop_co-auth_2}
    Suppose that, for every $i \in I$, and every $e \ni i$,
    $$\alpha_{i,e} \leq \gamma_i$$
    or
    $$\alpha_{i,e} \geq \lvert e \rvert \left(\beta_{i,e} (\mathfrak{m}+1) + \hat \beta_i (\mathfrak{m}_i-1)\right) + \gamma_i + 2 \kappa_i \mathfrak{m}_i.$$
    Then the linear congestion co-author game is fixed-sign componentwise monotonic.
    Consequently, it admits a trembling-hand equilibrium.
\end{proposition}

\begin{proof}
    See \ref{pf_prop_co-auth_2}.
\end{proof}

\paragraph{Numerical illustrations}

We now present three illustrative examples. The first one emphasizes the impact of budget constraints on the set of coalitionwise stable networks, while the other two illustrate the refinement properties of perturbed and trembling-hand equilibria.

\begin{example}
    This example illustrates how individual budget constraints may fundamentally affect the set of coalitionwise stable networks.
    Suppose that $I = \{1,2,3\}$, $E = \{e,f,g\}$, with $e = \{1,2\}$, $f = \{1,3\}$ and $g = \{1,2,3\}$, and $c_e = c_f = c_g = 1$.
    Moreover, assume that, for every $i \in I$, and every $h \in E$,
    $$\alpha_{i,h} = 2, \quad \beta_{i,h} = 0, \quad \gamma_i = 1, \quad \text{and} \quad \kappa_i = 0,$$
    so that
    $$u_1(x) = x_e + x_f + x_g, \quad u_2(x) = x_e + x_g, \quad u_3(x) = x_f + x_g,$$
    for every $x = (x_e,x_f,x_g) \in X(G^\mathrm{lc})$.
    Hence, every researcher strictly benefits from increasing the intensity of any project in which he participates.
    Suppose first that the budget constraints are nonbinding (i.e., $b_1 = b_2 = b_3 = +\infty$).
    Then every coalition intensity can be increased independently up to its capacity, so that the unique coalitionwise stable network is $(1,1,1)$.
    Assume now that $b_1 = b_2 = b_3 = 1$.
    Since researcher 1 belongs to every coalition, the budget constraint
    $$x_e + x_f + x_g \leq 1$$
    implies the remaining two constraints. Consequently,
    $$\mathcal{E}(G^\mathrm{lc}) = \{(x_e,x_f,x_g) : x_e + x_f + x_g = 1\}.$$
    Also, one can verify that
    $$\mathcal{P}(G^\mathrm{lc}) = \mathcal{T}(G^\mathrm{lc}) = \mathcal{E}(G^\mathrm{lc}).$$
    This example highlights the role played by budget constraints.
    Without resource scarcity, every project is developed at its maximal intensity, leading to a unique coalitionwise stable network.
    In contrast, when researchers face a unit budget, coalitionwise stability only requires researcher 1 to exhaust his available resources.
    The linear congestion co-author game therefore admits a continuum of coalitionwise stable networks, corresponding to the different ways in which the central researcher may allocate his effort between the two bilateral collaborations and the collective project. $\Diamond$
\end{example}

\begin{example}
    This example illustrates that coalitionwise stable networks need not be robust to perturbations.
    Suppose that $I = \{1,2,3\}$, $E = \{e,f\}$, with $e = \{1,2\}$ and $f = \{1,3\}$, $b_1 = b_2 = b_3 = +\infty$, and $c_e = c_f = 1$.
    Moreover, assume that, for every $i \in I$,
    $$\gamma_i = 1, \quad \text{and} \quad \kappa_i = 0,$$
    and that
    $$\alpha_{1,e} = 1, \quad \text{and} \quad \beta_{1,e} = 0,$$
    $$\alpha_{1,f} = 1, \quad \text{and} \quad \beta_{1,f} = \frac{1}{2},$$
    $$\alpha_{2,e} = 2, \quad \text{and} \quad \beta_{2,e} = 0,$$
    $$\alpha_{3,f} = 2, \quad \text{and} \quad \beta_{3,f} = 0,$$
    so that
    $$u_1(x) = -\frac{1}{2}x_ex_f - x_f^2, \quad u_2(x) = x_e, \quad u_3(x) = x_f,$$
    for every $x = (x_e,x_f) \in X(G^\mathrm{lc})$.
    When project $f$ is inactive, the intrinsic value of project $e$ exactly offsets researcher 1's participation cost.
    Researcher 1 is therefore indifferent to the intensity of project $e$.
    As soon as project $f$ is carried out with positive intensity, however, project $e$ generates a negative congestion effect for researcher 1, who then strictly prefers to reduce its intensity.
    Researchers 2 and 3, on the other hand, always benefit from increasing the intensity of their respective projects.
    A direct computation shows that
    $$\mathcal{E}(G^\mathrm{lc}) = \{(x_e,0) : x_e \in [0,1]\},$$
    whereas
    $$\mathcal{P}(G^\mathrm{lc}) = \{(0,0)\}.$$
    The economic interpretation of this refinement is particularly transparent.
    Every network of the form $(x_e,0)$ is coalitionwise stable, but positive values of $x_e$ are sustained only by a fragile indifference on the part of researcher 1.
    Such networks may be interpreted as inherited collaborations that researcher 1 has no strict incentive to terminate, even though he would not strictly benefit from creating or expanding them.
    However, this indifference disappears in every perturbed environment.
    Indeed, since $x_f^n \geq \varepsilon_f^n > 0$,
    $$\frac{\partial u_1}{\partial x_e}(x_e,x_f^n) = -\frac{1}{2}x_f^n < 0.$$
    Hence, the unilateral-decrease condition requires
    $$x_e^n = \varepsilon_e^n.$$
    Similarly, since $x_f^n > 0$,
    $$\frac{\partial u_1}{\partial x_f}(x_e^n,x_f) = -\frac{1}{2}x_e^n - 2x_f < 0.$$
    The same condition therefore implies
    $$x_f^n = \varepsilon_f^n.$$
    Consequently,
    $$x^n=(\varepsilon_e^n,\varepsilon_f^n) \longrightarrow (0,0).$$
    Thus, $(0,0)$ is the only network that does not rely on an exact indifference or on the persistence of an inherited collaboration. In this sense, the perturbed equilibrium concept selects the economically robust network and eliminates the fragile coalitionwise stable outcomes. $\Diamond$
\end{example}

\begin{example}
    This example illustrates that trembling-hand equilibria need not
    exist, even though perturbed equilibria do.
    Suppose that $I = \{1,2,3\}$, $E = \{e,f\}$, with $e = \{1,2\}$ and $f = \{1,3\}$, $b_1 = b_2 = b_3 = +\infty$, and $c_e = c_f = 1$.
    Moreover, assume that, for every $i \in I$,
    $$\gamma_i = 1, \quad \text{and} \quad \kappa_i = 0,$$
    and that
    $$\alpha_{1,e} = 2, \quad \text{and} \quad \beta_{1,e} = \frac{1}{2},$$
    $$\alpha_{1,f} = 1, \quad \text{and} \quad \beta_{1,f} = 0,$$
    $$\alpha_{2,e} = 2, \quad \text{and} \quad \beta_{2,e} = \frac{1}{2},$$
    $$\alpha_{3,f} = \frac{1}{2}, \quad \text{and} \quad \beta_{3,f} = 0,$$
    so that
    $$u_1(x) = u_2(x) = x_e - x_e^2-\frac{1}{2}x_ex_f, \quad u_3(x) = -\frac{1}{2}x_f,$$
    for every $x = (x_e,x_f) \in X(G^\mathrm{lc})$.
    Researchers 1 and 2 benefit from collaborating on project $e$,
    but congestion gradually reduces its marginal value.
    At the same time, researcher 3 always prefers to reduce the
    intensity of project $f$, which acts as a competing activity for
    researcher 1.
    One can verify that
    $$\mathcal{P}(G^\mathrm{lc}) \neq \emptyset,$$
    whereas
    $$\mathcal{T}(G^\mathrm{lc}) = \emptyset.$$
    This example highlights the stronger robustness requirement of the trembling-hand concept. $\Diamond$
\end{example}

\subsubsection{The Attention Dilution Co-author Game}

This second specification is closer in spirit to the original co-author model of Jackson and Wolinsky \cite{jackson_wolinsky_1996}.
Instead of assuming that congestion reduces productivity linearly, it assumes that the attention of each researcher is diluted across all the projects in which he is involved.
Thus, the contribution of a researcher to a given project decreases as his total workload increases.

Suppose that, for every $e \in E$, and every $i \in e$,
$$v_{i,e}: (s_j)_{j \in e} \longmapsto \alpha_{i,e} + \sum_{j \in e}\frac{\beta_{i,e,j}}{1+s_j} + \frac{\theta_{i,e}}{\prod_{j \in e}(1+s_j)},$$
where $\alpha_{i,e} \geq 0$ represents the intrinsic scientific value of project $e$ for researcher $i$, $\beta_{i,e,j} \geq 0$ measures the individual contribution of researcher $j$ to project $e$, and $\theta_{i,e} \geq 0$ captures a synergy effect among all members of the project.
Moreover, similarly as in the previous specification, suppose that, for every $i \in I$,
$$\psi_i: s \longmapsto \gamma_i s+\kappa_i s^2,$$
where $\gamma_i \geq 0$ and $\kappa_i \geq 0$.
The resulting utility for researcher $i \in I$ is therefore:
$$u_i(x) = \sum_{e \ni i} x_e \left(\alpha_{i,e} + \sum_{j \in e} \frac{\beta_{i,e,j}}{1+s_j(x)} + \frac{\theta_{i,e}}{\prod_{j \in e}(1+s_j(x))}\right) - \gamma_i s_i(x) - \kappa_i s_i(x)^2.$$
We refer to this specification as the \emph{attention dilution co-author game}, and denote this constrained network game by $G^\mathrm{ad}$.

This specification also satisfies the assumptions of Theorem \ref{thm_3}, under a suitable assumption on parameters.
For every $i \in I$, define
$$\mathfrak{m}_i = \lvert \{e \in E : i \in e\} \rvert, \quad \hat \beta_i = \max_{\substack{e \ni i\\j \in e}} \beta_{i,e,j}, \quad \text{and} \quad \hat \theta_i = \max_{e \ni i} \theta_{i,e}.$$

\begin{proposition} \label{prop_co-auth_3}
    Suppose that, for every $i \in I$ and every $e \ni i$,
    $$\kappa_i \geq \lvert e \rvert \mathfrak{m}_i \left(\hat \beta_i + \frac{\hat \theta_i}{2}(
    \lvert e \rvert + 1)\right).$$
    Then the attention dilution co-author game is continuous and componentwise quasiconcave.
    Consequently, it admits a perturbed equilibrium, and therefore an coalitionwise stable network.
\end{proposition}

\begin{proof}
    See \ref{pf_prop_co-auth_3}.
\end{proof}

As before, we provide a sufficient condition under which the same specification satisfies the monotonicity assumption used in Theorem \ref{thm_5}.

\begin{proposition} \label{prop_co-auth_4}
    Suppose that, for every $i \in I$, and every $e \ni i$,
    $$\alpha_{i,e} \leq \gamma_i - \hat \beta_i \lvert e \rvert - \theta_{i,e}$$
    or
    $$\alpha_{i,e} \geq \lvert e \rvert \left(\hat \beta_i \mathfrak{m}_i + \theta_{i,e} + \hat \theta_i (\mathfrak{m}_i - 1)\right) + \gamma_i + 2 \kappa_i \mathfrak{m}_i.$$
    Then the attention dilution co-author game is fixed-sign componentwise monotonic.
    Consequently, it admits a trembling-hand equilibrium.
\end{proposition}

\begin{proof}
    See \ref{pf_prop_co-auth_4}.
\end{proof}

\subsection{The Two-way Flow Game}

The \emph{two-way flow model} introduced by Bala and Goyal \cite{bala_goyal_2000} is one of the seminal models of strategic network formation.
In this framework, agents establish costly bilateral communication links in order to access information held by other agents.
The fundamental insight of the model is that information propagates through the network: a direct communication link provides access not only to the information of neighboring agents, but also to the information transmitted through indirect communication paths. Consequently, the value of a communication link depends on the entire network structure rather than only on the adjacent agents.
Bich and Morhaim \cite{bich_morhaim_2020} extend this framework to weighted networks by allowing communication links to have different intensities. Instead of considering only the existence of communication paths, they associate to each path the product of the intensities of its constituent links, thereby obtaining a continuous measure of information transmission and establishing an existence result for pairwise stable weighted networks.
Although this weighted formulation considerably enriches the original model, it still assumes that communication takes place exclusively through bilateral relationships. In many applications, however, information is naturally exchanged within groups rather than pairs of agents (e.g., research teams, collaborative platforms, industrial partnerships).
Furthermore, maintaining communication channels requires time, attention, computational resources, or organizational effort, implying that agents cannot participate intensively in every potential communication channel.
The constrained weighted network framework developed in this paper naturally incorporates these features by allowing communication channels to be coalitional while introducing both coalition capacities and individual resource constraints.

\begin{figure}[htbp]
    \centering
    \begin{tikzpicture}[
        >=Latex,
        agent/.style={
            circle,
            draw,
            thick,
            fill=white,
            minimum size=7mm,
            inner sep=0pt,
            font=\small
        },
        communication/.style={
            ->,
            very thick
        },
        every node/.style={font=\small}
    ]

        \node[agent] (i1) at (0,0) {1};
        \node[agent] (i2) at (2,1) {2};
        \node[agent] (i3) at (4,0) {3};
        \node[agent] (i4) at (6,0) {4};
        \node[agent] (i5) at (8,0) {5};

        \begin{scope}[on background layer]
            \draw[
                thick
            ]
                (2,0)
                ellipse[
                    x radius=2.75,
                    y radius=1.75
                ];

            \draw[
                thick,
                dashed
            ]
                (5,0)
                ellipse[
                    x radius=2,
                    y radius=1
                ];

            \draw[
                thick,
                dash dot
            ]
                (7,0)
                ellipse[
                    x radius=2,
                    y radius=1
                ];
        \end{scope}

        \node at (2,2.25)
            {$e_1=\{1,2,3\}$};

        \node at (5,1.5)
            {$e_2=\{3,4\}$};

        \node at (7,1.5)
            {$e_3=\{4,5\}$};

        \draw[communication]
            (i1.east)
            --
            (i3.west);

        \draw[communication,dashed]
            (i3.east)
            --
            (i4.west);

        \draw[communication,dash dot]
            (i4.east)
            --
            (i5.west);

    \end{tikzpicture}

    \caption{
        A coalitional communication path.
        Information held by agent 1 can reach agent 5 through the
        sequence of overlapping communication channels
        $(e_1,e_2,e_3)$.
        Agents 3 and 4 relay information between consecutive
        coalitions, and the intensity of the path is the product
        $x_{e_1}x_{e_2}x_{e_3}$.
    }
    \label{fig_3}
\end{figure}
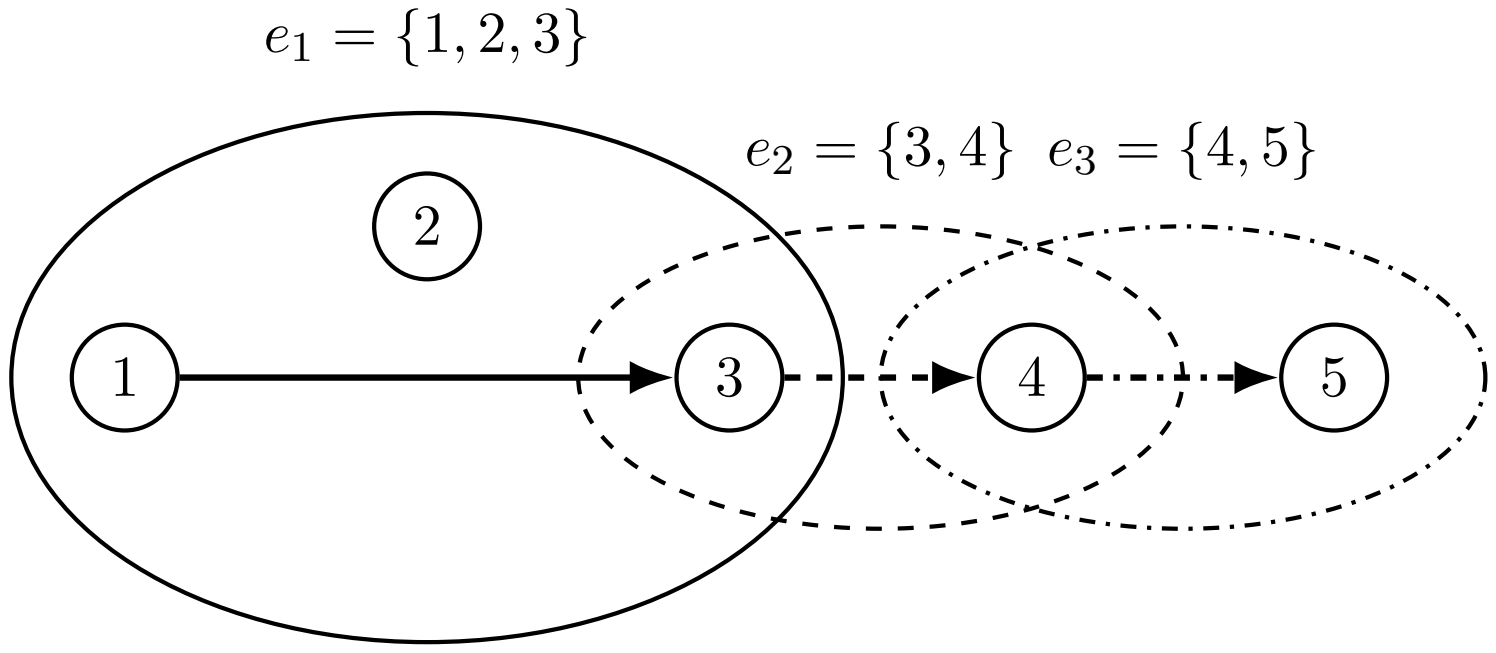

Consider a nonempty finite set $I$ of agents, and a nonempty set $E \subset 2^I \setminus \{\emptyset\}$ of potential communication channels, a budget vector
$(b_i)_{i \in I}$ and a capacity vector $(c_e)_{e \in E}$.
For every channel $e \in E$, the variable $x_e \in [0,c_e]$ denotes the intensity of communication within coalition $e$.
Each agent $i \in I$ faces the communication budget $b_i$, so that feasible communication profiles satisfy
$$\sum_{e \ni i} x_e \leq b_i.$$
For any $i \in I$, and any $j \neq i$, a \emph{coalitional communication path between $i$ and $j$} is a finite sequence $\pi = (e_1,\dots,e_N)$ of distinct coalitions such that
$$i \in e_1, \quad j \in e_N, \quad \text{and} \quad
e_n \cap e_{n+1} \neq \emptyset,
\quad
\forall n = 1,\dots,N-1.$$
The interpretation is that information can be relayed from one coalition to the next through their common members.
Let $\Pi_{ij}$ be the set of all communication paths between $i$ and $j$, for any $i \in I$, and any $j \neq i$, and let $\Pi = \bigcup_{\substack{k \in I\\\ell \neq k}} \Pi_{k\ell}$ be the \emph{set of coalitional communication paths}.
Figure \ref{fig_3} illustrates a coalitional
communication path.
For any coalitional communication path $\pi = (e_1,\dots,e_N) \in \Pi$, and any network $x$, the \emph{intensity of $\pi$ given $x$} is defined by
$$\nu(\pi \vert x) = \prod_{n = 1}^N x_{e_n}.$$
Following the weighted formulation of Bich and Morhaim \cite{bich_morhaim_2020}, define, for each $i \in I$,
$$v_i(x) = \sum_{j\neq i} \sum_{\pi \in \Pi_{ij}} \nu(\pi \vert x).$$
The utility of agent $i \in I$ is then given by
$$u_i(x) = v_i(x) - \kappa_i \sum_{e \ni i} x_e,$$
where $\kappa_i > 0$ denotes the marginal communication cost of agent $i$.
We refer to this constrained network game as the \emph{two-way flow game}, and denote it by $G^\mathrm{twf}$.

Similarly to the linear congestion co-author game, the two-way flow game satisfies the assumptions of Theorem \ref{thm_3}.

\begin{proposition} \label{prop_two-way_1}
    The two-way flow game is continuous and componentwise quasiconcave.
    Consequently, it admits a perturbed equilibrium, and therefore an coalitionwise stable network.
\end{proposition}

\begin{proof}
    Observe that, for any $i \in I$, and any $x \in X(G^\mathrm{twf})$, exactly as in the weighted two-way flow model of Bich and Morhaim \cite{bich_morhaim_2020}, the information value $v_i(x)$ of $x$ for agent $i$ is obtained as a finite sum of products of coalition intensities, which implies that $v_i$ is a multivariate polynomial (and therefore continuous).
    Moreover, for any $e \in E$, any $x_{-e} \in X(G^\mathrm{twf})_{-e}$, and any $i \in e$, notice that the function $x_e \mapsto v_i(x_e,x_{-e})$ is affine (since every coalition appears at most once in a coalitional communication path).
    This implies that $G^\mathrm{twf}$ is continuous and componentwise quasiconcave.
    Hence, by Theorem \ref{thm_3}, $G^\mathrm{twf}$ admits a perturbed equilibrium, and therefore an coalitionwise stable network.
\end{proof}

Unlike the extended co-author model, however, the marginal value of a communication channel depends on the entire collection of communication paths containing this channel. It therefore depends simultaneously on the topology of the communication network and on the intensities of all the other coalitions belonging to these paths. Consequently, no simple local sufficient condition guaranteeing fixed-sign componentwise monotonicity appears to exist. This observation highlights the practical relevance of Theorem \ref{thm_3}: even in network formation models where monotonicity is difficult to verify because of global interaction effects, the existence of perturbed equilibria still follows from the much milder assumptions of continuity and componentwise quasiconcavity.

\subsection{The Public-Good Provision Game}

A large literature studies strategic interactions on exogenously given networks (see, e.g., \cite{ballester_calvo-armengol_zenou_2006,bramoulle_kranton_d'amours_2014,galeotti_goyal_jackson_vega-redondo_yariv_2010}).
In these models, the network determines the pattern of strategic interactions, while agents choose their actions taking the interaction structure as fixed.
The framework developed in this paper allows one to reverse this perspective by endogenizing the network itself.
Consider for example the \emph{public-good provision model} introduced by Bramoullé and Kranton \cite{bramoulle_kranton_2007}.
In their framework, agents choose simultaneously a level of effort
in order to provide a local public good, given a fixed network of
bilateral relationships through which these efforts generate positive
externalities.
The utility of an agent depends on both his own contribution
and the contributions of his neighbors.
Our framework adopts a complementary perspective.
Rather than choosing effort levels endogenously, we assume that the
individual actions are fixed exogenously, while the coalitional interaction
structure itself is formed strategically.
This viewpoint is natural in situations where agents have already
chosen their individual level of investment, production or effort,
and subsequently decide how intensively to cooperate or exchange these
resources with others.
Moreover, unlike the original model, interactions are no longer
restricted to bilateral links.
Public goods may be produced or shared within arbitrary coalitions,
communication channels may exhibit heterogeneous intensities, and both
coalition capacities and individual budget constraints naturally limit
the amount of cooperation that can be sustained.

Consider a nonempty finite set $I$ of agents, a nonempty set $E \subset 2^I \setminus \{\emptyset\}$ of potential coalitions, a budget vector
$(b_i)_{i \in I}$ and a capacity vector $(c_e)_{e \in E}$.
Moreover, fix an action profile $(a_i)_{i \in I}$, where, for every $i \in I$, $a_i \geq 0$.
For every coalition $e \in E$, the variable $x_e \in [0,c_e]$ represents the intensity with which the members of coalition $e$ share the benefits generated by their individual efforts.
Each agent $i \in I$ faces the resource constraint $b_i$, so that feasible profiles satisfy
$$\sum_{e \ni i} x_e \leq b_i.$$
The utility of agent $i \in I$ is defined by
$$u_i(x) = v_i(a_i + \sum_{e \ni i} (x_e \sum_{j \in e \setminus \{i\}} a_j)) - \psi_i(a_i),$$
where
\begin{itemize}
\item $v_i: \mathbb {R}_+ \to \mathbb{R}$ is the benefit derived by agent $i$ from the total amount of public good available to him;
\item $\psi_i: \mathbb{R}_+ \to \mathbb{R}$ is the cost for agent $i$ associated with the exogenous action $a_i$.
\end{itemize}
We refer to this constrained network game as the \emph{public-good provision game}.

The public-good provision game provides a particularly simple illustration of our existence results.
Indeed, the utility of each agent $i \in I$ is obtained by composing the benefit function $v_i$ with an affine function of the coalition intensities.
Consequently, continuity, quasiconcavity and monotonicity are immediately inherited from the benefit functions.
The proofs are therefore straightforward and omitted.

\begin{proposition} \label{prop_pub-good_1}
    Suppose that, for every $i \in I$, $v_i$ is continuous and quasiconcave.
    Then the public-good provision game is continuous and componentwise quasiconcave.
    Consequently, it admits a perturbed equilibrium, and therefore an coalitionwise stable network.
\end{proposition}

\begin{proposition} \label{prop_pub-good_2}
    Suppose that, for every $i \in I$, $v_i$ is continuous and monotonic.
    Then the public-good provision game is fixed-sign componentwise monotonic.
    Consequently, it admits a trembling-hand equilibrium.
\end{proposition}

\section{Further Remarks}

The results obtained in this paper show that the introduction of feasibility constraints modifies the structure of weighted network formation problems.

At the level of stability, we have shown that the existence theorem of Bich and Morhaim \cite{bich_morhaim_2020} extends naturally to constrained environments.
Although budget and capacity restrictions introduce additional feasibility considerations, coalitionwise stable networks continue to exist under the same continuity and componentwise quasiconcavity assumptions.
This suggests that coalitionwise stability provides a robust extension of pairwise stability to constrained weighted network formation.

The analysis of perfection reveals a more substantial difference with the unconstrained framework.
In classical weighted network games, the trembling-hand and perturbative approaches introduced by Bich and Teteryatnikova \cite{bich_teteryatnikova_2023} are equivalent.
In constrained environments, however, feasibility restrictions alter the geometry of admissible deviations and make perturbations of the surrounding network significantly more delicate.
As illustrated by Propositions \ref{prop_1} and \ref{prop_2}, natural extensions of trembling-hand perfection may fail to exist, even in simple linear environments.
The notion of trembling-hand equilibrium introduced in this paper overcomes these difficulties by evaluating deviations relative to the feasible deviation set of the limit network.
Combined with the perturbative approach, this provides a framework in which existence can again be recovered.
In this sense, our results suggest that robustness considerations remain meaningful in constrained network formation, although they require a careful reformulation of the underlying perturbation concepts.

Several directions for future research naturally emerge from our analysis.
First, it would be interesting to investigate whether the monotonicity assumption used in our existence results can be weakened.
Second, one may study stronger notions of stability allowing simultaneous modifications of several coalitions, in the spirit of strong stability of Jackson--van den Nouweland \cite{jackson_van-den-nouweland_2005}.
More generally, constrained weighted network formation provides a natural bridge between network formation theory, matching theory, and resource allocation problems. We hope that the concepts introduced in this paper may contribute to a deeper understanding of the role played by feasibility constraints in the emergence and stability of complex interaction structures.

\appendix

\section{Proof of Theorem \ref{thm_1}} \label{pf_thm_1}

The proof follows the spirit of Theorem 6.1 in Bich--Morhaim \cite{bich_morhaim_2017}.
Namely, consider the correspondence $\Psi: \mathcal{F}(G) \to 2^{\mathcal{F}(G)}$ defined as follows.
For any feasible networks $x,y \in \mathcal{F}(G)$, $y \in \Psi(x)$ if and only if there exists a coalition $e \in E$ such that one of the following conditions holds:
\begin{enumerate}
	\item There exists $d_e \in [0,x_e)$ such that $y = (d_e,x_{-e})$ and
    $$\exists i \in e, u_i(y) > u_i(x).$$
	\item There exists $d_e \in (x_e,c_e]$ such that $y = (d_e,x_{-e})$ and
    $$\forall i \in e, u_i(y) > u_i(x).$$
\end{enumerate}
We also recall the following result due to Yannelis and Prabhakar (see Theorem 5.2 in \cite{yannelis_prabhakar_1983}).
	
\begin{theorem*}[Yannelis--Prabhakar \cite{yannelis_prabhakar_1983}] \label{thm_yannelis_prabhakar}
	Let $K$ be a nonempty, compact and convex subset of some finite-dimensional Euclidean space.
	Moreover, let $\Phi: K \to 2^K$ be a lower hemicontinuous correspondence such that, for every $x \in K$, $x \notin \mathrm{Conv}(\Phi(x))$.
	Then there exists $\bar x \in K$ such that $\Phi(\bar x) = \emptyset$.
\end{theorem*}

Since $\mathcal{F}(G)$ is nonempty, compact and convex, it remains to show that the correspondence $\Psi$ satisfies the assumptions of the Yannelis--Prabhakar theorem.
The main difficulty, compared to the proof in Bich--Morhaim \cite{bich_morhaim_2017}, is to show that the correspondence $\Psi$ is lower hemicontinuous.

\subsection{$\Psi$ is lower hemicontinuous}

Since $\mathcal{F}(G)$ is a subset of a finite-dimensional Euclidean space, recall that the correspondence $\Psi$ is lower hemicontinuous if and only if for every $x \in \mathcal{F}(G)$, every sequence $(x^n)$ in $\mathcal{F}(G)$ converging to $x$, and every $y \in \Psi(x)$, there exists a subsequence $(x^{n_m})$ and a sequence $(y^m)$ in $\mathcal{F}(G)$ such that $y^m \longrightarrow y$ and $y^m \in \Psi(x^{n_m})$, for all $m$.
Fix $x \in \mathcal{F}(G)$, a sequence $(x^n)$ in $\mathcal{F}(G)$ converging to $x$, and $y \in \Psi(x)$.
By definition of $\Psi$, there exists a coalition $e \in E$ such that either condition $1$ or condition $2$ holds.
Fix such a coalition.

\subsubsection{Case one: there exists a profitable unilateral decrease}

Assume condition $1$ holds.
Then there exists $d_e \in [0,x_e)$ such that $y = (d_e,x_{-e})$ and
$$\exists i \in e, u_i(y) > u_i(x).$$
Since $d_e < x_e$, there exists $\bar n_1$ such that $d_e < x^n_e$, for all $n \geq \bar n_1$.
By continuity of $u_i$, there exists $\bar n_2$ such that $u_i(d_e,x^n_{-e}) > u_i(x^n)$, for all $n \geq \bar n_2$.
Let $\bar n = \max \{\bar n_1,\bar n_2\}$, and define, for any $m$, $x^{n_m} = x^{\bar n+m}$ and
$$y^m = (d_e,x^{n_m}_{-e}),$$
and simply observe that, by construction, $y^m$ is feasible and belongs to $\Psi(x^{n_m})$.

\subsubsection{Case two: there exists a profitable multilateral increase}

Assume condition $2$ holds.
Then there exists $d_e \in (x_e,c_e]$ such that $y = (d_e,x_{-e})$ and
$$\forall i \in e, u_i(y) > u_i(x).$$
Define
$$I_e^< = \left\{i \in e : \sum_{f \ni i} y_f < b_i\right\} \quad \text{and} \quad I_e^= = \left\{i \in e : \sum_{f \ni i} y_f = b_i\right\}$$
(if one of these two sets is empty, the argument simplifies and we omit it).
For any $n$, define
$$d^n_e = \min \left\{d_e,\min \left\{b_i - \sum_{\substack{f \ni i\\f \neq e}} x^n_f : i \in I_e^=\right\}\right\}.$$
Since $x_e < d_e$, there exists $\bar n_1$ such that
\begin{equation} \label{eq_1}
    x^n_e < d_e,
\end{equation}
for all $n \geq \bar n_1$.
For every $i \in e$, by continuity of $u_i$, there exists $\bar n^i_2$ such that
\begin{equation} \label{eq_2}
    u_i(d^n_e,x^n_{-e}) > u_i(x^n),
\end{equation}
for all $n \geq \bar n^i_2$.
For every $i \in I_e^<$, since $\sum_{f \ni i} y_f < b_i$, there exists $\bar n_3^i$ such that
\begin{equation} \label{eq_3}
    \sum_{\substack{f \ni i\\f \neq e}} x^n_f + d^n_e < b_i,
\end{equation}
for all $n \geq \bar n_3^i$.
For every $i \in I_e^=$, since $\sum_{f \ni i} y_f = b_i$ and $x_e < d_e$, we have that $\sum_{f \ni i} x_f < b_i$, which implies that there exists $\bar n_3^i$ such that
\begin{equation} \label{eq_4}
    x^n_e < b_i - \sum_{\substack{f \ni i\\f \neq e}} x^n_f,
\end{equation}
for all $n \geq \bar n_3^i$.
Let $\bar n = \max \{\bar n_1,\max \{\bar n_2^i : i \in e\},\max \{\bar n_3^i : i \in e\}\}$, and define, for any $m$, $x^{n_m} = x^{\bar n+m}$, $d_e^{n_m} = d_e^{\bar n+m}$ and
$$y^m = (d^{n_m}_e,x^{n_m}_{-e}).$$
We now check that $y^m$ belongs to $\Psi(x^{n_m})$, for all $m$:
\begin{enumerate}
    \item First, we show that $(y^m)$ is a sequence in $\mathcal{F}(G)$:
    \begin{enumerate}
        \item Since $d_e \leq c_e$, we directly have that
        $$y^m_e = d^{n_m}_e = \min \left\{d_e,\min \left\{b_i - \sum_{\substack{f \ni i\\f \neq e}} x^{n_m}_f : i \in I_e^=\right\}\right\} \leq c_e$$
        (also, $y^m_f = x^{n_m}_f \leq c_f$, for any $f \neq e$).
        \item Let $i \in I$:
        \begin{enumerate}
            \item If $i \in I_e^<$, then, by (\ref{eq_3}), we directly obtain that $\sum_{f \ni i} y^m_f < b_i$.
            \item If $i \in I_e^=$, then, since $\sum_{\substack{f \ni i\\f \neq e}} x^{n_m}_f + (b_i - \sum_{\substack{f \ni i\\f \neq e}} x^{n_m}_f) = b_i$, we have that
            $$\sum_{\substack{f \ni i\\f \neq e}} x^{n_m}_f + \min \left\{d_e,\min \left\{b_j - \sum_{\substack{f \ni j\\f \neq e}} x^{n_m}_f : j \in I_e^=\right\}\right\} = \sum_{f \ni i} y^m_f \leq b_i.$$
            \item If $i \notin e$, then feasibility follows directly, by definition of $(x^{n_m})$.
        \end{enumerate}
    \end{enumerate}
    \item Finally, we show that $y^m \in \Psi(x^{n_m})$. Consider the coalition $e$ and the weight $d^{n_m}_e$:
    \begin{enumerate}
        \item By (\ref{eq_1}) and by (\ref{eq_4}), we obtain that
        $$y^m_e = d^{n_m}_e = \min \left\{d_e,\min \left\{b_i - \sum_{\substack{f \ni i\\f \neq e}} x^n_f : i \in I_e^=\right\}\right\} > x^{n_m}_e.$$
        \item By (\ref{eq_2}), we obtain that, for every $i \in e$,
        $$u_i(y^m) = u_i(d^{n_m}_e,x^{n_m}_{-e}) > u_i(x^{n_m}).$$
    \end{enumerate}
    Hence, $y^m$ satisfies condition $2$, which implies that $y^m \in \Psi(x^{n_m})$.
\end{enumerate}
This proves that $\Psi$ is lower hemicontinuous.

\subsection{For every $x \in \mathcal{F}(G)$, $x \notin \mathrm{Conv}(\Psi(x))$}

Let $x \in \mathcal{F}(G)$, and assume by contradiction that $x \in \mathrm{Conv}(\Psi(x))$.
Then, there exists $(x^\ell)_{\ell \in L} \subset \Psi(x)$ and weights $(\theta^\ell)_{\ell \in L}$ such that
$$x = \sum_{\ell \in L} \theta^\ell x^\ell,$$
for some (finite) index set $L$.
Fix a coalition $e \in E$ such that $x^\ell_e \neq x_e$, for some $\ell$.
Let $L_e = \{\ell \in L : x^\ell_e \neq x_e\}$.
Then $x_e$ is a convex combination of $(x^\ell_e)_{\ell \in L_e}$.
Note that there exists $\bar \ell_1 \in L_e$ such that $(x^{\bar \ell_1}_e,x_{-e})$ satisfies condition $1$, and $\bar \ell_2 \in L_e$ such that $(x^{\bar \ell_2}_e,x_{-e})$ satisfies condition $2$.
Let $i \in e$ such that $u_i(x^{\bar \ell_1}_e,x_{-e}) > u_i(x)$.
Note that $u_i(x^{\bar \ell_2}_e,x_{-e}) > u_i(x)$.
Since $x_e$ lies strictly between $x^{\bar \ell_1}_e$ and $x^{\bar \ell_2}_e$, quasiconcavity of $u_i$ implies that
$$u_i(x) \geq \min \left\{u_i(x^{\bar \ell_1}_e,x_{-e}),u_i(x^{\bar \ell_2}_e,x_{-e})\right\} > u_i(x),$$
a contradiction.
Hence, $x \notin \mathrm{Conv}(\Psi(x))$.

\subsection{Application of the Yannelis--Prabhakar theorem}
		
From the two last steps, all the assumptions of the Yannelis--Prabhakar theorem are satisfied.
Therefore, there exists $\bar x \in \mathcal{F}(G)$ such that $\Psi(\bar x) = \emptyset$.
This implies exactly the two stability conditions for $\bar x$. $\square$

\section{Proof of Theorem \ref{thm_3}} \label{pf_thm_3}

Let $(\varepsilon^n,\eta^n)$ be a sequence in $X(G) \times [0,+\infty)^I$ that converges to $(\mathbf{0},\mathbf{0})$ and that satisfies, for any $n$, the following conditions:
\begin{enumerate}
    \item For every $i \in I$, $0 < \eta^n_i < b_i$.
    \item For every $e \in E$, $0 < \varepsilon^n_e < \frac{c_e}{2}$.
    \item $\mathcal{F}^n(G)$ is nonempty.
\end{enumerate}
(Since, by assumption, $\mathrm{int}(\mathcal{F}(G))$ is nonempty, such a sequence exists.)
For any $n$, consider the constrained network game
$$G^n = (I,E,(b_i^n)_{i \in I},(c_e^n)_{e \in E},(u_i^n)_{i \in I}),$$
where, for every $i \in I$,
$$b^n_i = b_i - \eta_i^n - \sum_{e \ni i} \varepsilon_e^n$$
and
$$u^n_i: x \in X(G^n) \longmapsto u_i(x+\varepsilon^n) \in \mathbb{R},$$
and where, for every $e \in E$,
$$c_e^n = c_e - 2\varepsilon_e^n.$$
Note that, for every $n$, the game $G^n$ is well defined (in particular, for every $x \in X(G^n)$, $x+\varepsilon^n$ belongs to $X(G)$, so that, for every $i \in I$, $u^n_i$ is well defined), and observe that it is also continuous and componentwise quasiconcave.
Hence, by Theorem \ref{thm_1}, $G^n$ admits an coalitionwise stable network
$\bar x^n$, forall $n$.
Now, for every $n$, define
$$\bar y^n = \bar x^n + \varepsilon^n \in \mathcal{F}(G).$$
Since $\mathcal{F}(G)$ is compact, we can assume, up to a subsequence, that $(\bar y^n)$ converges to some $\bar y \in \mathcal{F}(G)$.
We now show that the sequence $(\varepsilon^n,\eta^n,\bar y^n)$ satisfies the required properties, which will imply that $\bar y$ is a perturbed equilibrium of $G$.

\subsection{For any $n$, $\bar y^n$ belongs to $\mathcal{F}^n(G)$}

Fix $n$.
We have that, for every $e \in E$, $\bar x^n_e$ belongs to $[0,c_e - 2\varepsilon_e^n]$, which implies that $\bar y^n = \bar x^n + \varepsilon^n$ belongs to $[\varepsilon_e^n,c_e - \varepsilon_e^n]$.
On the other hand, for every $i \in I$,
$$\sum_{e \ni i} \bar y^n_e = \sum_{e \ni i} \bar x^n_e + \sum_{e \ni i} \varepsilon^n_e \leq b^n_i + \sum_{e \ni i} \varepsilon^n_e = b_i - \eta^n_i.$$
Therefore, $\bar y^n \in \mathcal{F}^n(G)$.

\subsection{Unilateral decrease condition}

Fix $n$, $e \in E$, $d^n_e \in [\varepsilon^n_e,\bar y^n_e)$ and $i \in e$.
Define
$$\tilde d^n_e = d^n_e - \varepsilon^n_e,$$
which belongs to $[0,\bar x^n_e)$.
By definition, we have that
$$u_i(d^n_e,\bar y^n_{-e}) = u^n_i(\tilde d^n_e,\bar x^n_{-e}),$$
and since $\bar x^n \in \mathcal{E}(G^n)$, we obtain that
$$u^n_i(\tilde d^n_e,\bar x^n_{-e}) \leq u^n_i(\bar x^n),$$
where
$$u^n_i(\bar x^n) = u_i(\bar y^n).$$
Therefore, the unilateral decrease condition is verified.

\subsection{Multilateral increase condition}

Fix $n$, $e \in E$ and $d_e^n \in (\bar y_e^n,\mu_e^n(G,\bar y_{-e}^n)]$.
Define
$$\tilde d^n_e = d^n_e - \varepsilon^n_e,$$
which belongs to $(\bar x^n_e,\mu_e(G^n,\bar x^n_{-e})]$, where we recall that
\begin{align*}
    \mu_e(G^n,\bar x^n_{-e}) &= \min \left\{c^n_e,\min_{i \in e} \left(b^n_i - \sum_{\substack{f \ni i\\f \neq e}} \bar x^n_f\right)\right\}\\
    &= \min \left\{c_e - 2\varepsilon^n_e,\min_{i \in e} \left(b_i - \eta_i^n - \sum_{f \ni i} \varepsilon_f^n - \sum_{\substack{f \ni i\\f \neq e}} \bar x^n_f\right)\right\}.
\end{align*}
Indeed, we have that
$$d^n_e \leq \mu_e^n(G,\bar y_{-e}^n) = \min \left\{c_e-\varepsilon^n_e,\min_{i \in e} \left(b_i - \eta^n_i - \sum_{\substack{f \ni i\\f \neq e}} \bar y^n_f\right)\right\},$$
which implies that
\begin{align*}
    \tilde d^n_e &= d^n_e - \varepsilon^n_e\\
    &\leq \min \left\{c_e-\varepsilon^n_e,\min_{i \in e} \left(b_i - \eta^n_i - \sum_{\substack{f \ni i\\f \neq e}} \bar y^n_f\right)\right\} - \varepsilon^n_e\\
    &= \min \left\{c_e-\varepsilon^n_e,\min_{i \in e} \left(b_i - \eta^n_i - \sum_{\substack{f \ni i\\f \neq e}} (\bar x^n_f + \varepsilon^n_f)\right)\right\} - \varepsilon^n_e\\
    &= \min \left\{c_e - 2\varepsilon^n_e,\min_{i \in e} \left(b_i - \eta_i^n - \sum_{f \ni i} \varepsilon_f^n - \sum_{\substack{f \ni i\\f \neq e}} \bar x^n_f\right)\right\}\\
    &= \mu_e(G^n,\bar x^n_{-e}).
\end{align*}
By definition, we have that, for every $i \in I$,
$$u_i(d^n_e,\bar y^n_{-e}) = u^n_i(\tilde d^n_e,\bar x^n_{-e}),$$
and since $\bar x^n \in \mathcal{E}(G^n)$, we obtain that there exists $i \in e$ such that
$$u^n_i(\tilde d^n_e,\bar x^n_{-e}) \leq u^n_i(\bar x^n),$$
where
$$u^n_i(\bar x^n) = u_i(\bar y^n).$$
Therefore, the multilateral increase condition is verified. $\square$

\section{Proof of Theorem \ref{thm_4}} \label{pf_thm_4}

The proof is similar to that of Theorem 3 in Bich--Teteryatnikova \cite{bich_teteryatnikova_2023}.
The only substantial difference is that the presence of budget constraints requires the simultaneous construction of a sequence $(\eta^n)$ of budget perturbations, in addition to the sequence $(\varepsilon^n)$ of capacity perturbations.
We state first the following lemma.

\begin{lemma}
    For every $e \in E$, and every $i \in e$, exactly one of the following two conditions holds:
    \begin{enumerate}
        \item For every $x_{-e} \in X(G)_{-e}$, $u_i(\cdot,x_{-e})$ is strictly decreasing.
        \item For every $x_{-e} \in X(G)_{-e}$, $u_i(\cdot,x_{-e})$ is strictly increasing.
    \end{enumerate}
\end{lemma}

\begin{proof}
    Recall that, by assumption, for every $x_{-e} \in X(G)_{-e}$, $u_i(\cdot,x_{-e})$ is strictly monotonic.
    Define
    $$\varphi_i: x_{-e} \in X(G)_{-e} \longmapsto u_i(c_e,x_{-e})-u_i(0,x_{-e}) \in \mathbb{R}.$$
    Since $u_i$ is continuous, the function $\varphi_i$ is also continuous.
    Moreover, for every $x_{-e} \in X(G)_{-e}$, because $u_i(\cdot,x_{-e})$ is strictly monotonic, we have that $\varphi_i(x_{-e}) \neq 0$.
    Therefore, since $X(G)_{-e}$ is connected, $\varphi_i(X(G)_{-e})$ is a connected subset of $\mathbb{R} \setminus \{0\}$.
    Consequently, either
    $$\varphi_i(x_{-e}) < 0, \quad \forall x_{-e} \in X(G)_{-e},$$
    implying that $u_i(\cdot,x_{-e})$ is strictly decreasing, for every $x_{-e} \in X(G)_{-e}$, or
    $$\varphi_i(x_{-e}) > 0, \quad \forall x_{-e} \in X(G)_{-e}.$$
    implying that $u_i(\cdot,x_{-e})$ is strictly increasing, for every $x_{-e} \in X(G)_{-e}$.
\end{proof}

Also, for every $e \in E$, define
$$I_e^{\downarrow} = \{i \in e : \forall x_{-e} \in X(G)_{-e}, u_i(\cdot,x_{-e}) \text{ is strictly decreasing}\}$$
and
$$I_e^{\uparrow} = \{i \in e : \forall x_{-e} \in X(G)_{-e}, u_i(\cdot,x_{-e}) \text{ is strictly increasing}\}.$$
By the previous lemma, we have that
$$e = I_e^{\downarrow} \sqcup I_e^{\uparrow}.$$

\subsection{If a network is a perturbed equilibrium of $G$, then it is a trembling-hand equilibrium of $G$}

Let $\bar x \in \mathcal{P}(G)$ be a perturbed equilibrium of $G$, and let $(\varepsilon^n,\eta^n,x^n)$ be an associated perturbing sequence.
Observe that, for any $n$, and every $e \in E$, we necessarily have that
$$x^n_e =
\left\{
\begin{array}{ccc}
    \varepsilon^n_e&\text{if}&I_e^{\downarrow} \neq \emptyset\\
    \mu_e^n(G,x^n_{-e})&\text{else}&
\end{array}
\right.,$$
and passing to the limit yields
$$\bar x_e =
\left\{
\begin{array}{ccc}
    0&\text{if}&I_e^{\downarrow} \neq \emptyset\\
    \mu_e(\bar x_{-e})&\text{else}&
\end{array}
\right..$$
Now, choose any sequence $(\tilde x^n)$ in $\mathcal{F}(G)$ that converges to $\bar x$ such that, for every $n$, $\tilde x^n \in \mathrm{int}(\mathcal{F}(G))$.
Fix $e \in E$.
If $I_e^{\downarrow} \neq \emptyset$, then, for any $n$, unilateral decrease condition is vacuously satisfied, and any agent $i \in I_e^{\downarrow}$ is such that $u_i(d_e,\tilde x^n_{-e}) < u_i(\bar x_e,\tilde x^n_{-e})$, for any $d_e \in (\bar x_e,\mu_e(G,\bar x_{-e})]$.
If $e = I_e^{\uparrow}$, then, for any $n$, multilateral increase condition is vacuously satisfied, and every agent $i \in e$ is such that $u_i(d_e,\tilde x^n_{-e}) < u_i(\bar x_e,\tilde x^n_{-e})$, for any $d_e \in [0,\bar x_e)$.
Hence $\bar x$ is a trembling-hand equilibrium of $G$.

\subsection{If a network is a trembling-hand equilibrium of $G$, then it is a perturbed equilibrium of $G$}

Let $\bar x$ be a trembling-hand equilibrium of $G$, and let $(x^n)$ be an associated sequence.
Also, consider a sequence $(\delta^n)$ in $[0,1]$ such that $\delta^n \longrightarrow 0$, and $\delta^n > 0$, for all $n$.
Similarly as before, note that, for every $e \in E$,
$$\bar x_e =
\left\{
\begin{array}{ccc}
    0&\text{if}&I_e^{\downarrow} \neq \emptyset\\
    \mu_e(G,\bar x_{-e})&\text{else}&
\end{array}
\right..$$
First, define a sequence $(\varepsilon^n)$ in $X(G)$ as follows.
For every $n$, and every $e \in E$, set
$$\varepsilon^n_e =
\left\{
\begin{array}{ccc}
    x^n_e&\text{if}&I_e^{\downarrow} \neq \emptyset\\
    c_e-x^n_e&\text{if}&e = I_e^{\uparrow} \text{ and } \bar x_e = c_e\\
    \min\{x^n_e,c_e-x^n_e,\delta^n\}&\text{if}&e = I_e^{\uparrow} \text{ and } \bar x_e < c_e
\end{array}
\right..$$
Because $x^n \longrightarrow \bar x$ and $\delta^n \longrightarrow 0$, we have that $\varepsilon^n \longrightarrow \mathbf{0}$.
In particular, this means that there exists $\bar n_1$ such that, for every $n \geq \bar n_1$, and every $e \in E$,
$$0 < \varepsilon_e^n < \frac{c_e}{2}.$$
Second, define a sequence $(\eta^n)$ in $[0,+\infty)^I$ as follows.
For every $n$, and every $i \in I$, set
$$\eta^n_i =
\left\{
\begin{array}{ccc}
    b_i - \sum_{e \ni i} x^n_e&\text{if}&\sum_{e \ni i} \bar x_e = b_i\\
    \min\{b_i - \sum_{e \ni i} x^n_e,\delta^n\}&\text{if}&\sum_{e \ni i} \bar x_e < b_i
\end{array}
\right..$$
Because $x^n \longrightarrow \bar x$ and $\delta^n \longrightarrow 0$, we have that $\eta^n \longrightarrow \mathbf{0}$.
In particular, this means that there exists $\bar n_2$ such that, for every $n \geq \bar n_2$, and every $i \in I$,
$$0 < \eta^n_i < b_i.$$
Finally, set $\bar n = \max \{\bar n_1,\bar n_2\}$, and define the sequence
$$(\tilde x^m,\tilde \varepsilon^m,\tilde \eta^m) = (x^{\bar n+m},\varepsilon^{\bar n+m},\eta^{\bar n+m}).$$
By construction, for any $m$, $\tilde x^m$ belongs to
$$\mathcal{\widetilde F}^m(G) = \left\{z^m \in \prod_{e \in E} [\tilde \varepsilon^m_e,c_e-\tilde \varepsilon^m_e] : \forall i \in I, \sum_{e \ni i} z^m_e \leq b_i - \tilde \eta^m_i\right\}.$$
It remains to show that unilateral decrease and multilateral increase conditions hold in order to prove that $\bar x$ is a perturbed equilibrium of $G$.
Fix $e \in E$ and $m$.

\subsubsection{Unilateral decrease condition}

Suppose first that $I_e^{\downarrow}$ is nonempty.
In that case, we have that $\tilde \varepsilon^m_e = \tilde x^m_e$, which implies that unilateral decrease condition is vacuously verified.
Suppose now that $e = I_e^{\uparrow}$, and let $d^m_e \in [\tilde \varepsilon^m_e,\tilde x^m_e)$ and $i \in e$.
For every $i \in e$, because $u_i(\cdot,\tilde x^m_{-e})$ is strictly increasing, we directly obtain that $u_i(d^m_e,\tilde x^m_{-e}) < u_i(\tilde x^m)$.

\subsubsection{Multilateral increase condition}

Suppose first that $I_e^{\downarrow}$ is nonempty, and let $d^m_e \in (\tilde x^m_e,\mu_e^m(G,\tilde x^m_{-e})]$.
For every $i \in I_e^{\downarrow}$, because $u_i(\cdot,\tilde x^m_{-e})$ is strictly decreasing, we directly obtain that $u_i(d^m_e,\tilde x^m_{-e}) < u_i(\tilde x^m)$.
Suppose now that $e = I_e^{\uparrow}$, and let $d^m_e \in (\tilde x^m_e,\mu_e^m(G,\tilde x^m_{-e})]$.
We show that
$$\mu_e^m(G,\tilde x^m_{-e}) = \tilde x^m_e,$$
which will imply that multilateral increase condition is vacuously verified.
If $\bar x_e = c_e$, then, by definition of $\tilde \varepsilon^m_e$,
$$\tilde x^m_e = c_e - \tilde \varepsilon^m_e.$$
Therefore,
$$\mu_e^m(G,\tilde x^m_{-e}) = \min \left\{c_e-\tilde \varepsilon^m_e,\min_{i \in e} \left(b_i - \tilde \eta^m_i - \sum_{\substack{f \ni i\\f \neq e}} \tilde x^m_f\right)\right\} \leq c_e - \tilde \varepsilon^m_e = \tilde x^m_e.$$
On the other hand, since $\tilde x^m$ belongs to $\mathcal{\widetilde F}^m(G)$, we also have that 
$$\tilde x^m_e \leq \mu_e^m(G,\tilde x^m_{-e}),$$
Thus,
$$\mu_e^m(G,\tilde x^m_{-e}) = \tilde x^m_e.$$
Suppose now that $\bar x_e < c_e$, and recall that, since $I_e^{\downarrow}$ is empty, we have that
$$\bar x_e = \mu_e(G,\bar x_{-e}) = \min \left\{c_e,\min_{i \in e} \left(b_i - \sum_{\substack{f \ni i\\f \neq e}} \bar x_f\right)\right\}.$$
Hence, we have that there exists $i \in e$ such that $\bar x_e = b_i - \sum_{\substack{f \ni i\\f \neq e}} \bar x_f$, i.e.,
$$b_i = \sum_{f \ni i} \bar x_f.$$
By definition of $\tilde \eta^m_i$, we obtain that
$$\tilde x^m_e = b_i - \tilde \eta^m_i - \sum_{\substack{f \ni i\\f \neq e}} \tilde x^m_f.$$
Therefore,
$$\mu_e^m(G,\tilde x^m_{-e}) = \min \left\{c_e-\tilde \varepsilon^m_e,\min_{j \in e} \left(b_j - \tilde \eta^m_j - \sum_{\substack{f \ni j\\f \neq e}} \tilde x^m_f\right)\right\} \leq b_i - \tilde \eta^m_i - \sum_{\substack{f \ni i\\f \neq e}} \tilde x^m_f = \tilde x^m_e.$$
Again, since $\tilde x^m$ belongs to $\mathcal{\widetilde F}^m(G)$, we also have that 
$$\tilde x^m_e \leq \mu_e^m(G,\tilde x^m_{-e}),$$
Thus,
$$\mu_e^m(G,\tilde x^m_{-e}) = \tilde x^m_e. \quad \square$$

\section{Proof of Theorem \ref{thm_5}} \label{pf_thm_5}

By the fixed-sign componentwise monotonicity assumption, for every $e \in E$, and every $i \in e$, consider $\sigma_{e,i} \in \{-1,1\}$ such that the function $x_e \mapsto \sigma_{e,i}u_i(x_e,x_{-e})$ is nondecreasing, for every $x_{-e} \in X(G)_{-e}$.
Now, for every $n$, consider the constrained network game
$$G^n = (I,E,(b_i)_{i \in I},(c_e)_{e \in E},(u^n_i)_{i \in I}),$$
where, for every $i \in I$,
$$u^n_i: x \in X(G) \longmapsto u_i(x) + \frac{1}{n} \sum_{e \ni i} \sigma_{e,i}x_e,$$
and observe that $G^n$ is continuous.
We also have the following result.

\begin{lemma}
    For any $n$, $G^n$ is componentwise strictly monotonic.
\end{lemma}

\begin{proof}
    Consider $e \in E$, $x_{-e} \in X(G)_{-e}$, and $i \in e$.
    Moreover, let $\omega_e,\omega'_e \in [0,c_e]$ such that $\omega_e < \omega'_e$.
    We have that
    $$u^n_i(\omega'_e,x_{-e}) - u^n_i(\omega_e,x_{-e}) = (u_i(\omega'_e,x_{-e}) - u_i(\omega_e,x_{-e})) + \frac{1}{n} \sigma_{e,i} (\omega'_e - \omega_e).$$
    If $\sigma_{e,i} = -1$, then, by definition, $u_i(\cdot,x_{-e})$ is nonincreasing, which implies that $u_i(\omega'_e,x_{-e}) - u_i(\omega_e,x_{-e}) \leq 0$.
    Thus,
    $$u^n_i(\omega'_e,x_{-e}) - u^n_i(\omega_e,x_{-e}) \leq -\frac{1}{n} (\omega'_e - \omega_e) < 0,$$
    i.e., $u^n_i(\cdot,x_{-e})$ is strictly decreasing.
    If $\sigma_{e,i} = 1$, then, by definition, $u_i(\cdot,x_{-e})$ is nondecreasing, which implies that $u_i(\omega'_e,x_{-e}) - u_i(\omega_e,x_{-e}) \geq 0$.
    Thus,
    $$u^n_i(\omega'_e,x_{-e}) - u^n_i(\omega_e,x_{-e}) \geq \frac{1}{n} (\omega'_e - \omega_e) > 0,$$
    i.e., $u^n_i(\cdot,x_{-e})$ is strictly increasing.
\end{proof}

For every $n$, because $G^n$ is both continuous and componentwise strictly monotonic, Corollary \ref{coro_1} implies that $G^n$ admits a trembling-hand equilibrium $\bar x^n \in \mathcal{T}(G^n) \subset \mathcal{F}(G)$.
Since $\mathcal{F}(G)$ is compact, consider a subsequence $(\bar x^{n_m})$ that converges to some $\bar x \in \mathcal{F}(G)$.
On the other hand, for every $e \in E$, define
$$I_e^- = \{i \in e : \sigma_{e,i} = -1\}$$
and
$$I_e^+ = \{i \in e : \sigma_{e,i} = 1\}.$$
Also, note that, for every $e \in E$,
$$e = I_e^- \sqcup I_e^+.$$

\begin{lemma} \label{lem_1}
    For every $e \in E$, if $I_e^-$ is nonempty, then $\bar x_e = 0$.
\end{lemma}

\begin{proof}
    Simply observe that, for every $i \in I_e^-$, and every $m$, because $u_i^{n_m}(\cdot,\bar x^{n_m}_{-e})$ is strictly decreasing and $\bar x^{n_m}$ is a trembling-hand equilibrium of $G^{n_m}$, we must have that $\bar x_e^{n_m} = 0$.
    Passing to the limit yields $\bar x_e = 0$.
\end{proof}

\begin{lemma} \label{lem_2}
    For every $e \in E$, if $I_e^-$ is empty, then $\bar x_e = \mu_e(G,\bar x_{-e})$.
\end{lemma}

\begin{proof}
    By definition, since $I_e^- = \emptyset$, we have that $e = I_e^+$.
    Then, for every $i \in e$, and every $m$, the function $u_i^{n_m}(\cdot,\bar x^{n_m}_{-e})$ is strictly increasing.
    Since $\bar x^{n_m}$ is a trembling-hand equilibrium of $G^{n_m}$, we must have that $\bar x_e^{n_m} = \mu_e(G^{n_m},\bar x^{n_m}_{-e}) = \mu_e(G,\bar x^{n_m}_{-e})$.
    Finally, because $x_{-e} \mapsto \mu_e(G,x_{-e})$ is continuous, passing to the limit yields $\bar x_e = \mu_e(G,\bar x_{-e})$.
\end{proof}

Now, fix $z \in \mathrm{int}(\mathcal{F}(G))$, consider a sequence $(\delta^n)$ in $[0,1]$ such that $\delta^n \longrightarrow 0$, and $0 < \delta^n < 1$, for all $n$.
Also, define, for every $n$,
$$T^n: x \in X(G) \longmapsto (1-\delta^n) x + \delta^n z \in X(G).$$
We show that $\bar x$ is a trembling-hand equilibrium of $G$, using the sequence
$$(z^n) = (T^n(\bar x))$$
in $\mathcal{F}(G)$.
Since $\mathcal{F}(G)$ is convex and $z \in \mathrm{int}(\mathcal{F}(G))$, note that, for every $n$, $z^n$ also belongs to $\mathrm{int}(\mathcal{F}(G))$.
Fix $e \in E$.
Suppose first that $I_e^-$ is nonempty.
Because, by Lemma \ref{lem_1}, $\bar x_e = 0$, unilateral decrease condition is vacuously verified.
For any $i \in I_e^-$, and every $n$, we have that $u_i(\cdot,z^n_{-e})$ is nonincreasing, which directly implies that multilateral increase condition is also verified.
Finally, suppose that $e = I_e^+$.
Because, by Lemma \ref{lem_2}, $\bar x_e = \mu_e(G,\bar x_{-e})$, multilateral increase condition is vacuously verified.
For any $i \in e$, and every $n$, we have that $u_i(\cdot,z^n_{-e})$ is nondecreasing, which directly implies that unilateral decrease condition is also verified. $\square$

\section{Proofs in the Extended Co-author Model}

\subsection{Proofs in the Linear Congestion Co-author Game}

Observe first that, for every $e \in E$, every $i \in e$, and every $x_{-e} \in X(G^\mathrm{lc})_{-e}$,
\begin{align*}
    &u_i(\cdot,x_{-e})'(x_e) = \alpha_{i,e} - \beta_{i,e}\left(\sum_{j \in e} s_j(x_e,x_{-e}) + \lvert e \rvert x_e\right) - \sum_{\substack{f \ni i\\f \neq e}} \beta_{i,f} \lvert f\cap e \rvert x_f\\
    &- \gamma_i - 2\kappa_i s_i(x_e,x_{-e})
\end{align*}
and
$$u_i(\cdot,x_{-e})''(x_e) = -2\beta_{i,e} \lvert e \rvert - 2\kappa_i,$$
for all $x_e \in [0,c_e]$.
Moreover, recall that, for every $i \in I$,
$$\mathfrak{m}_i = \lvert \{e \in E : i \in e\} \rvert \quad \text{and} \quad \hat \beta_i = \max_{e \ni i} \beta_{i,e},$$
and that
$$\mathfrak{m} = \max_{j \in I} \mathfrak{m}_j.$$

\subsubsection{Proof of Proposition \ref{prop_co-auth_1}} \label{pf_prop_co-auth_1}

Continuity is immediate.
Moreover, simply observe that, for every $e \in E$, every $i \in e$, every $x_{-e} \in X(G^\mathrm{lc})_{-e}$, and every $x_e \in [0,c_e]$, $u_i(\cdot,x_{-e})''(x_e) \leq 0$.
Hence, for every $e \in E$, every $i \in e$, and every $x_{-e} \in X(G^\mathrm{lc})_{-e}$, $u_i(\cdot,x_{-e})$ is concave, therefore quasiconcave.
The conclusion follows from Theorem \ref{thm_3}. $\square$

\subsubsection{Proof of Proposition \ref{prop_co-auth_2}} \label{pf_prop_co-auth_2}

Fix $e \in E$, $i \in e$, and $x_{-e} \in X(G^\mathrm{lc})_{-e}$.
If
$$\alpha_{i,e} \leq \gamma_i,$$
then we directly have that $u_i(\cdot,x_{-e})'(x_e) \leq 0$, for all $x_e \in [0,c_e]$, so that $u_i(\cdot,x_{-e})$ is nonincreasing.
Assume now that
\begin{equation} \label{eq_9}
    \alpha_{i,e} \geq \lvert e \rvert \left(\beta_{i,e} (\mathfrak{m}+1) + \hat \beta_i (\mathfrak{m}_i-1)\right) + \gamma_i + 2 \kappa_i \mathfrak{m}_i.
\end{equation}
Fix $x_e \in [0,c_e]$ and recall that, by definition,
\begin{equation} \label{eq_10}
    x_f \leq c_f \leq 1 \quad (f \in E),
\end{equation}
so that, in particular,
$$s_j(x_e,x_{-e}) \leq \mathfrak{m}_j \leq \mathfrak{m} \quad (j \in I).$$
Together, (\ref{eq_9}) and (\ref{eq_10}) thus imply that $u_i(\cdot,x_{-e})'(x_e) \geq 0$, so that $u_i(\cdot,x_{-e})$ is nondecreasing.
Hence the game is fixed-sign componentwise monotonic.
The existence of a trembling-hand equilibrium follows from Theorem \ref{thm_5}. $\square$

\subsection{Proofs in the Attention Dilution Co-author Game}

For every $i \in I$, define
$$r_i(x)=\frac{1}{1+s_i(x)} \quad (x \in X(G^\mathrm{ad})),$$
and, for every $e \in E$, define
$$R_e(x)=\prod_{j \in e} r_j(x) \quad (x \in X(G^\mathrm{ad})).$$
Then, for every $i \in I$,
{\small
$$u_i(x) = \sum_{e \ni i} x_e \left(\alpha_{i,e} + \sum_{j\in e} \beta_{i,e,j} r_j(x) + \theta_{i,e} R_e(x)\right) - \gamma_i s_i(x) - \kappa_i s_i(x)^2 \quad (x \in X(G^\mathrm{ad})).$$
}
Now, observe that, for every $e \in E$, every $i \in e$, and every $x_{-e} \in X(G^\mathrm{ad})_{-e}$,
\begin{align*}
    &u_i(\cdot,x_{-e})'(x_e) = \alpha_{i,e} + \sum_{j \in e} \beta_{i,e,j} r_j(x_e,x_{-e}) + \theta_{i,e} R_e(x_e,x_{-e})\\
    &- x_e \left(\sum_{j \in e} \beta_{i,e,j} r_j(x_e,x_{-e})^2 + \theta_{i,e} R_e(x_e,x_{-e}) \sum_{j\in e} r_j(x_e,x_{-e}) \right)\\
    &- \sum_{\substack{f \ni i\\f \neq e}} x_f \left(\sum_{j \in f \cap e}\beta_{i,f,j} r_j(x_e,x_{-e})^2 + \theta_{i,f} R_f(x_e,x_{-e}) \sum_{j \in f \cap e} r_j(x_e,x_{-e})\right)\\
    &- \gamma_i - 2\kappa_i s_i(x_e,x_{-e})
\end{align*}
and
{\scriptsize
\begin{align*}
    &u_i(\cdot,x_{-e})''(x_e) = - 2\left(\sum_{j \in e} \beta_{i,e,j}r_j(x_e,x_{-e})^2 + \theta_{i,e} R_e(x_e,x_{-e}) \sum_{j \in e} r_j(x_e,x_{-e})\right)\\
    &+ x_e \left(2\sum_{j \in e} \beta_{i,e,j} r_j(x_e,x_{-e})^3 + \theta_{i,e} R_e(x_e,x_{-e}) \left[\left(\sum_{j \in e} r_j(x_e,x_{-e})\right)^2 + \sum_{j \in e} r_j(x_e,x_{-e})^2\right]\right)\\
    &+ \sum_{\substack{f \ni i\\f \neq e}} x_f \left(2\sum_{j \in f \cap e} \beta_{i,f,j} r_j(x_e,x_{-e})^3 + \theta_{i,f} R_f(x_e,x_{-e}) \left[\left(\sum_{j \in f \cap e} r_j(x_e,x_{-e}) \right)^2 + \sum_{j \in f \cap e} r_j(x_e,x_{-e})^2\right]\right)\\
    &- 2\kappa_i,
\end{align*}
}
for all $x_e \in [0,c_e]$.
Moreover, recall that, for every $i \in I$,
$$\mathfrak{m}_i = \lvert \{e \in E : i \in e\} \rvert, \quad \hat \beta_i = \max_{\substack{e \ni i\\j \in e}} \beta_{i,e,j}, \quad \text{and} \quad \hat \theta_i = \max_{e \ni i} \theta_{i,e}.$$

\subsubsection{Proof of Proposition \ref{prop_co-auth_3}} \label{pf_prop_co-auth_3}

Continuity is immediate.
Fix $e \in E$, $i \in e$, and $x_{-e} \in X(G^\mathrm{ad})_{-e}$, and suppose that
\begin{equation} \label{eq_11}
    \kappa_i \geq \lvert e \rvert \mathfrak{m}_i \left(\hat \beta_i + \frac{\hat \theta_i}{2}(
    \lvert e \rvert + 1)\right).
\end{equation}
Moreover, fix $x_e \in [0,c_e]$ and recall that, by definition,
\begin{equation} \label{eq_12}
    x_f \leq c_f \leq 1 \quad (f \in E),
\end{equation}
so that, in particular,
$$r_j(x_e,x_{-e}) \leq 1 \quad (j \in I)$$
and
$$R_f(x_e,x_{-e}) \leq 1 \quad (f \in E),$$
implying that
{\footnotesize
\begin{align*}
    &x_e \left(2\sum_{j \in e} \beta_{i,e,j} r_j(x_e,x_{-e})^3 + \theta_{i,e} R_e(x_e,x_{-e}) \left[\left(\sum_{j \in e} r_j(x_e,x_{-e})\right)^2 + \sum_{j \in e} r_j(x_e,x_{-e})^2\right]\right)\\
    &\leq 2 \hat \beta_i \lvert e \rvert + \hat \theta_i(
    \lvert e \rvert^2 + \lvert e \rvert)
\end{align*}
}
and
{\scriptsize
\begin{align*}
    &\sum_{\substack{f \ni i\\f \neq e}} x_f \left(2\sum_{j \in f \cap e} \beta_{i,f,j} r_j(x_e,x_{-e})^3 + \theta_{i,f} R_f(x_e,x_{-e}) \left[\left(\sum_{j \in f \cap e} r_j(x_e,x_{-e}) \right)^2 + \sum_{j \in f \cap e} r_j(x_e,x_{-e})^2\right]\right)\\
    &\leq (\mathfrak{m}_i - 1) \left(2 \hat \beta_i \lvert e \rvert + \hat \theta_i(
    \lvert e \rvert^2 + \lvert e \rvert)\right).
\end{align*}
}
Together, (\ref{eq_11}) and (\ref{eq_12}) thus imply that $u_i(\cdot,x_{-e})''(x_e) \leq 0$, so that $u_i(\cdot,x_{-e})$ is concave, therefore quasiconcave.
The conclusion follows from Theorem \ref{thm_3}. $\square$

\subsubsection{Proof of Proposition \ref{prop_co-auth_4}} \label{pf_prop_co-auth_4}

Fix $e \in E$, $i \in e$, and $x_{-e} \in X(G^\mathrm{ad})_{-e}$.
Suppose first that
\begin{equation} \label{eq_13}
    \alpha_{i,e} \leq \gamma_i - \hat \beta_i \lvert e \rvert - \theta_{i,e},
\end{equation}
and observe that (\ref{eq_12}) and (\ref{eq_13}) together imply that $u_i(\cdot,x_{-e})'(x_e) \leq 0$, so that $u_i(\cdot,x_{-e})$ is nonincreasing.
Suppose now that
\begin{equation} \label{eq_14}
    \alpha_{i,e} \geq \lvert e \rvert \left(\hat \beta_i \mathfrak{m}_i + \theta_{i,e} + \hat \theta_i (\mathfrak{m}_i - 1)\right) + \gamma_i + 2 \kappa_i \mathfrak{m}_i.
\end{equation}
and observe that (\ref{eq_12}) and (\ref{eq_14}) together imply that $u_i(\cdot,x_{-e})'(x_e) \geq 0$, so that $u_i(\cdot,x_{-e})$ is nondecreasing.
Hence the game is fixed-sign componentwise monotonic.
The existence of a trembling-hand equilibrium follows from Theorem \ref{thm_5}. $\square$

\end{document}